\documentclass[12pt, preprint]{aastex}
\newcommand{\cc}{\mbox{cm$^{-3}$}}

\newcommand{\nhtwo}{\mbox{n$_{H_{2}}$}}

\def\m17{M~17}               
\def\Htwo{H$_2$}               
\def\HtwoO{H$_2$O}             
\def\HCOp{HCO$^+$}             
\def\thCO{$^{13}$CO}           
\def\CeiO{C$^{18}$O}           
\def\Otwo{O$_2$}               
\def\NtwoHp{N$_2$H$^+$}        
\def\CHthreeOH{CH$_3$OH}       

\def\m{\ts {\rm m}}
\def\kms{\ts {\rm km\ts s$^{-1}$}}

\let\ts=\thinspace

\def\swash2o{$1_{10} - 1_{01}$}               
\def\isoh2o{2$_{12}-1_{01}$}               
\def\oHtwoO{o-H$_2$O}               
\def\G0{G$_0$}               

\begin{document}

\title{A Survey of 557 GHz Water Vapor Emission in the NGC 1333 Molecular Cloud}

\author{Edwin A. Bergin}
\affil{Harvard-Smithsonian Center for Astrophysics, 60 Garden Street,
Cambridge, MA 02138} \email{ebergin@cfa.harvard.edu}

\author{Michael J. Kaufman}
\affil{Department of Physics, San Jose State University, One Washington Square,
San Jos\'e, CA 95192-0106} 
\affil{NASA/Ames Research Center, MS 245-3, Moffett Field, CA 94035}
\email{kaufman@ism.arc.nasa.gov}

\author{Gary J. Melnick}
\affil{Harvard-Smithsonian Center for Astrophysics, 60 Garden Street,
Cambridge, MA 02138} \email{gmelnick@cfa.harvard.edu}

\author{Ronald L. Snell}
\affil{Department of Astronomy,
University of Massachusetts, Amherst, MA 01003}
\email{snell@astro.umass.edu}

\author{John E. Howe}
\affil{Department of Astronomy,
University of Massachusetts, Amherst, MA 01003}
\email{jhowe@astro.umass.edu}

\slugcomment{accepted by The Astrophysical Journal}
\begin{abstract}
Using NASA's Submillimeter Wave Astronomy Satellite (SWAS)
we have examined the production of water in quiescent and 
shocked molecular gas through a survey of 
the 556.936 GHz \swash2o\ transition of ortho-\HtwoO\ in the NGC 1333 molecular core.
These observations reveal broad emission lines associated with the IRAS~2, 
IRAS~4, IRAS~7, and HH7-11 outflows.  Towards 3 positions we detect narrow
($\Delta v \sim$ 2--3 km s$^{-1}$) emission lines clearly associated with the
ambient gas.   The SWAS observations, with a resolution of $\sim 4'$, 
are supplemented with observations from the Infrared Space Observatory (ISO) and
by an unbiased survey of a $\sim$17$'$ $\times$ 15$'$  area, 
with $\sim 50''$ resolution, in the low-J transitions of CO, \thCO , \CeiO ,
\NtwoHp , \CHthreeOH , and SiO.    

Using these combined data sets, with consistent
assumptions, we find beam-averaged ortho-\HtwoO\ abundances of $> 10^{-6}$ relative
to \Htwo\ for all four outflows.   A comparison of SWAS and ISO water data is
consistent with non-dissociative shock models, provided the majority of the
ortho-\HtwoO\ \swash2o\ emission arises from cool post-shock material with enhanced abundances.
In the ambient gas the ortho-\HtwoO\ abundance 
is found to lie between 0.1 -- 1 $\times 10^{-7}$ relative to \Htwo\ and is enhanced
when compared to cold pre-stellar molecular cores.  
A comparison of the water emission with tracers of dense condensations and
shock chemistry finds no clear correlation.  However, the water emission appears
to be associated with the presence of luminous external heating sources which
power the reflection nebula and the photodissociation (PDR) region.
Simple PDR models are capable of reproducing the water and high-J \thCO\ emission,
suggesting that a PDR may account for the excitation of water in low density
undepleted gas as suggested by Spaans \& van Dishoeck (2001).
\end{abstract}


\section{Introduction}

Water in cold quiescent molecular gas is theorized to form through a series 
of ion-molecule chemical reactions that occur early in the reaction sequence that
links atomic oxygen to H$_2$, the most abundant molecule. 
Recently NASA's Submillimeter-Wave Astronomy Satellite detected water vapor emission
arising from ortho-\HtwoO\ (\oHtwoO ) ground state transition (\swash2o ) in numerous dense 
cores within extended molecular clouds \citep{snell_h2o}.  
Interestingly, there appears to be a 
dichotomy in the presence of water detected by SWAS in the sense that emission 
from water vapor has yet to be detected in cold (T $\le 15-20$ K) star-less objects 
such as TMC-1, B68, or $\rho$ Oph D \citep{snell_h2o, bergin_b68h2o}.  
However, in
warmer (T $>$ 20 K) giant molecular cloud cores, 
which are associated with sites of multiple star
formation, water is readily detected.  In each case the derived water abundance,
or abundance limit, is well below theoretical predictions, which 
is believed to be the result of the freeze-out, or depletion, of oxygen onto grain
surfaces \citep{bergin_impl, viti_h2o, charnley_swasiso}.   
It is not clear if the non-detections in colder cores 
are the result of any intrinsic difference in the chemistry due to lower
temperature, because the excitation conditions for the \swash2o\ transition 
favors emission in warm gas.  
An alternative has been proposed by \citet{spaans_h2o} who
suggest that the enhanced penetration of ultraviolet (UV) photons
in clumpy giant molecular clouds lowers the water vapor column density 
through photodissociation but 
enhances the emissivity due to the heating of the gas by the UV radiation.
This would allow ion-molecule chemistry to account for water detections in
warm giant molecular clouds but
colder clouds, such as the star-less objects, require freeze-out of oxygen in the
form of \HtwoO\ \citep{spaans_h2o}. 

Conditions in the interstellar medium also favor water production through 
a variety of other mechanisms besides ion-molecule chemistry.  Water is believed to be
produced efficiently through reactions on the surfaces of cold dust grains 
\citep{tielens_hagen}. 
However, the sublimation temperature of water ice is $\sim 110$ K \citep{fraser_h2olab}, 
much greater
than typical dust grain temperatures, so this frozen reservoir is not returned
to the gas phase. 
An additional formation channel is linked to a series of neutral-neutral
reactions with activation barriers that can be overcome in gas that is
heated to temperatures greater than $\sim$300 K \citep{wagner_oxy, graff_oxy}.   
Such high temperatures can be found in shocked gas, where material entrained
in the bipolar outflows associated with the birth of a star impact the surrounding
gas, or in the immediate vicinity of embedded sources.
Indeed ESA's {\em Infrared Space Observatory} (ISO) and SWAS
have convincingly detected water vapor emission in hot gas towards several 
star-forming regions
\citep{liseau_hh54, cecc_h2o, harwit_h2o, melnick_bnkl,
neufeld_h2o, molinari_hh711, wright_h2o, 
giannini_h2o, maret_h2o, benedettini_h2o}.   
The water abundance determined for these 
regions is also significantly enhanced, by several orders of magnitude, over
that estimated in the ambient gas within other nearby clouds by \citet{snell_h2o}. 
These abundance enhancements 
are expected to persist for $\sim 10^{5}$ yr, whereupon the molecules 
deplete onto grain mantles \citep{bmn98}.  
This led to the suggestion by \citet{bmn98} that molecular gas could be 
chemically enriched in \HtwoO\ due to repeated exposure to numerous shocks.
Furthermore, once the dynamical
effects of a shock have dissipated, enhanced abundances of water vapor, above
that produced by ion-molecule reactions, might
be found as a shock ``legacy'' in the low velocity quiescent gas.

To examine the production of water in the interstellar medium we present
the results of a biased survey of the NGC 1333 molecular core for emission
in the 557 GHz ground state of ortho-\HtwoO .  
This source is relatively close, at a distance of 220 pc \citep{cernis90}, and
is an excellent laboratory to investigate the various potential mechanisms  
for water formation and emission.   
First, NGC 1333 is a well known reflection nebula illuminated by 
two nearby B stars, BD $+$30.549 (B8V) and SVS~3 (B6V).    The radiation
from these external sources allows for a search for water emission
associated with photodissociation regions (PDR) within the clumpy 
NGC 1333 cloud \citep{sk01},
as suggested by \citet{spaans_h2o}. 
Second, near-infrared photometry has revealed
a dense cluster of young stars 
embedded within the molecular gas \citep{asr94, lal96, as97}.  
These clustered young
stars have powered a burst of energetic outflow activity manifested
in the detection of clumps of vibrationally excited H$_2$ emission
and a  large number of Herbig-Haro objects \citep{asr94, bdr96, hl95}.
These clumps of high excitation 
are associated with a complex cluster of overlapping molecular outflows 
that dominate the structure in the dense core of 
the NGC 1333 molecular cloud and apparently have carved out large
cavities within the dense gas \citep{lcl96, lefloch_dust, ks00}. 
Thus, it is possible that the outflows have enriched the quiescent gas with water vapor,
which is also a focus of this paper.

We report detections of spectrally resolved water vapor emission
in several distinct outflows, and also from several positions
tracing the quiescent gas.  
To aid in the analysis of the SWAS data,
we present results from a supplementary survey 
of NGC 1333, using the Five College Radio Astronomy Observatory (FCRAO),
in transitions of CO, \thCO , 
\NtwoHp , \CHthreeOH , and SiO.  
Our analysis of these combined data sets suggests that the quiescent water abundance
is enhanced relative to cold pre-stellar molecular cores.  
Despite the pervasive presence of outflowing
gas we find no supporting evidence that the quiescent water vapor is due to past
shock episodes.  Rather the correspondence between the water emission
and high-J \thCO\ emission suggests that a PDR could provide additional heating to
excite water in relatively low density quiescent gas. 
The observations and data reduction
are presented in \S 2, and in \S 3 we present the results and a  
comparison of the integrated emissions maps with the water vapor distribution.
In \S 4 we derive an estimate of the \oHtwoO\ abundance in the high and low
velocity gas. In \S 5 we discuss the origin of water vapor emission in NGC 1333. 
Our conclusions are given in \S 6.

\section{Observations}

The SWAS \oHtwoO\ observations presented here were obtained during 
678 hours between 1999 January -- March. 1999, 1999 August,
2000 January -- February, 2000 August -- October, and 2002 January.  The spacecraft
was used in nod mode, involving alternately nodding the spacecraft to an off-source 
position free of emission ($\alpha_{off} =$ 03:28:07.9, $\delta_{off} =$
$+$33:10:22; J2000).  All survey data were obtained with the center
referenced to the position of SVS~13: $\alpha =$ 03:29:03.7, 
$\delta =$ 31:16:02.7 (J2000).  
Table~1 summarizes the specific positions surveyed,
the total integration time, and the rms in the 557 GHz band.
A total of 9 separate pointings were made, in some cases with the
elliptical SWAS beam slightly overlapped.    This survey is biased
in that known centers of activity were specifically centered in
the antenna beam (e.g. IRAS~2, 4, 7, and HH7-11).  Additional observations were
obtained offset from these targeted positions. 
Towards each position we have
assigned a SWAS survey number (SS\#; provided in Table~1) 
that is used for ease in reference in
cases where there is no clear association with an existing well known source.
Table~2 provides a listing of the
line frequencies, beam size, and adopted main beam efficiency.
All data were reduced with the standard SWAS
pipeline described by \citet{melnick_swas}.  
Towards each offset position, along with \oHtwoO ,
SWAS simultaneously observed transitions
of [C I] ($^3$P$_1$--$^3$P$_0$), \thCO\ (J = 5--4),
and \Otwo\ ($3_3$--$1_2$).  In this work we primarily use the 
\oHtwoO\ and \thCO\ data, with a velocity resolution of 1.0 \kms , and a sampling
of 0.6 \kms .    Molecular oxygen was not detected towards
any position down to a typical rms of $\sigma(T_A^*)$ = 42 mK (3$\sigma$).  
With all data in the survey averaged together there is still no 
\Otwo\ detection, with a 3$\sigma$ upper limit of 12 mK. 

Beam sampled maps of the mm-wave rotational transitions of $^{12}$CO, \thCO , \CeiO ,
\NtwoHp , \HCOp , and SiO were obtained with the FCRAO 14 m antenna
on 2000 November 12 and 13, and 2001 March 11 and 12.  
Pointing was checked frequently on the SiO masers (R Cas and Orion) with
an accuracy of $< 4''$.  Typical
SSB system temperatures were 500 K ($^{12}$CO), 290 K (\thCO ), 350 K (\CeiO ),
200 K (\NtwoHp ), 250 K (\HCOp ), and 200 K (SiO). 
Table~2 summarizes other relevant observational information.
The maps encompassed the entire region surveyed by SWAS and covered a spatial
region of nearly 17$'$ $\times$ 15$'$.  
The SWAS and FCRAO data are presented here on the
$T_A^*$ scale and are uncorrected for antenna efficiency. 
All analysis is performed correcting for main beam efficiency, provided in Table~2.

\section{Results}

\subsection{SWAS}

We have obtained a wealth of additional higher resolution data to supplement 
the SWAS observations.  These data are primarily used to place the SWAS
observations in context with the general structure of the core.  In this section
we present only the integrated intensity maps of \NtwoHp\ J = 1--0 (Figure~1) 
and \thCO\ J = 1--0  (Figure~2)
to delineate the location of dense condensations and low density gas in NGC 1333. 
Figure~1 also provides the 
spectra of \oHtwoO\ $1_{01} - 
1_{10}$ taken towards 9 positions in NGC 1333.  
The observed intensity, 
line center velocity, and velocity width of the \oHtwoO\ \swash2o\ detections are 
given in Tables~3 and 4.  Relevant parameters for additional molecular transitions
are also provided.
In Figure~1 the dashed line in each
spectra denotes the
velocity of the quiescent gas determined through Gaussian fits to the  
\NtwoHp\ emission convolved to a simulated 4$'$ circularized SWAS beam.  
The positions of the SWAS \oHtwoO\
observations, along with the elliptical SWAS beam given at the observed rotation
angle, are shown superposed on the integrated emission map of \NtwoHp\ (1--0).
\NtwoHp\ has been shown to be a good tracer of the dust distribution 
\citep{bergin_ic5146, tafalla_depletion}, and in this case shows an excellent correspondence
to the observed dust continuum emission (compare with \citet{lefloch_dust, sk01}; 
see also \citet{difrancesco_iras4}).

Figure 1 shows clear detections of broad ($\Delta v > 20$ \kms ) 
water vapor emission lines towards
IRAS~2, IRAS~4, and HH 7-11 (SVS~13).  
Given the narrowness of the line observed in the direction of SVS~12 there 
is no evidence for water emission in the SVS~12 outflow.  
In the direction of IRAS~7 we 
detect relatively broad emission, $\Delta v \sim 8.5$ \kms , consistent with 
an origin in the IRAS~7 outflow.
In each case where outflow emission is detected, at the velocity of the
ambient gas,
there is evidence for absorption from foreground
material.   
The coincidence of these observations with strong \NtwoHp\ emission
suggests that there is indeed quiescent gas towards each line of sight.
Water has previously been detected in IRAS~4 
by ISO and SWAS, and HH7-11 by ISO\footnote{The 
spectrally unresolved ISO detections of
high lying \HtwoO\ transitions can also be interpreted as arising from envelope 
gas in proximity to IRAS4a,b \citep{cecc_h2o}.  
However, the
spatially unresolved, but spectrally resolved, SWAS observations clearly
demonstrate that emission in the 557 ground state, at least, predominantly
arises within the outflow \citep{neufeld_h2o}.}
\citep{cecc_h2o, neufeld_h2o, molinari_hh711, giannini_h2o}.   
We note that there is some beam overlap in these observations, particularly
in the case of IRAS~2 and IRAS~7 with HH 7-11.  However,
for IRAS~2 and HH7-11 the \swash2o\ peak flux  is comparable (Figure~1),
which would not be the case if either detection
is the solely the result of contamination from the other.  
Thus, it is likely that SWAS has detected water emission from both flows. 
In the case of IRAS~7 it is possible there could be some contribution from
an extended HH7-11 flow.  As this is currently uncertain we assume
all the emission arises solely from IRAS~7.

For 3 positions, SS1, SS2-SVS~12, and SS5, there are significant detections
of a narrow, $\Delta v \sim$ 2 -- 3 \kms , component of \oHtwoO\ emission
associated  with cold dormant material.   In each case 
the emission is blue-shifted with respect to the systemic velocity as defined
by \CeiO\ (compare Gaussian fit line center velocities in Table~4) and \NtwoHp .   
There is no clear correlation of the ambient water emission to
the dense condensations traced in \NtwoHp , as only the SVS~12 (SS2) observation
coincides with a dense clump.  Indeed the water emission towards SS5 
lies inside the well known cavity between SVS~12 and SVS~13 \citep{warin_etal96,
lcl96, lefloch_dust, sk01}.  

In Figure~2 we provide the observed spectra of the J = 5--4 transition of \thCO , 
while Table~5 lists the integrated intensity and line parameters.
Similar to Fig.~1, we show the beam positions of the SWAS observations placed
on a map of the \thCO\ J = 1--0 integrated emission.
The J = 5--4 SWAS observations were obtained simultaneously 
with \HtwoO\ and have identical pointing.  
Clearly the strongest \thCO\ J = 5--4 emission, by
nearly a factor of two, is not associated with the more well known molecular
cores, such as HH7-11, IRAS~2, or even IRAS~4.   Rather this strong emission 
is seen towards SS1 and SS2-SVS12, which are close to the known
sources of reflection nebulosity, BD +30.549 and SVS~3.
These sources are likely heating the nearby gas in the PDR, which excites
high-J emission lines.
The SWAS observations towards these two positions also overlap the strongest clump
of \thCO\ J = 1--0 emission.
However, the \thCO\ J = 1--0 emission maximum does not correlate with 
with any dense condensation as traced
by \NtwoHp\ (compare Figures~1 and 2).

The 2 positions with strong J = 5--4 emission also coincide with 
2 detections  of quiescent \oHtwoO .  
However, quiescent \HtwoO\ is also seen towards SS5, which has moderate intensity 
\thCO\ J = 5--4 emission in comparison to other positions. 
The association of strong J = 5--4 emission and the presence of the nearby 
luminous sources, along with a partial correlation with detections of quiescent  \oHtwoO ,
suggests that the
presence of a photodissociation region (PDR) in the northern portions of NGC 1333
may be related to the presence of quiescent water emission.  
This question is investigated in \S 5.

\subsection{FCRAO}

Within the central portion of the NGC 1333 core lie a number of individual
objects and systems (SVS~13, IRAS~2, IRAS~4) and their associated 
outflows that have been the
subject of a multitude of molecular investigations \citep{snell_edwards, 
bc90, sandell_iras2, 
blake_iras4, warin_etal96, lcl96, lefloch_sio, codella_sio, difrancesco_iras4, 
rudolph_hh711}. 
However, these investigations generally focused on individual sources and most of the 
molecular transitions have not previously been
observed over a large scale in the unbiased fashion required for this study.
The detection of abundant SiO in the ambient gas 
in NGC 1333 \citep{lefloch_sio, codella_sio} 
has particular relevance to any \oHtwoO\ ambient emission. 
SiO is rarely detected in cold material
\citep{ziurys_sio, codella_sio}; rather it appears to be found predominantly in hot
shocked material \citep{bachiller_araa}.  The origin of SiO in shocks is believed
to originate from the release of silicon from grain cores by ablation
in shocks, followed by gas-phase chemical
reactions \citep{schilke_sio, caselli_sio, may_sputter}.  The quiescent SiO in NGC 1333
could originate from previous shock episodes or through current
interactions with the dense condensations \citep{lefloch_sio, codella_sio}. 
It is also well known that the
abundance of \CHthreeOH\ is enhanced in outflows \citep{bachiller_araa}, and
broad emission components have been detected in IRAS~2  and 4
\citep{sandell_iras2, blake_iras4}.
As such, a shock origin that created SiO in the ambient gas
could be operative for both \HtwoO\ and \CHthreeOH .  

In Figure~3 we show a sample of the spectra taken from all species and
transitions in the survey towards HH7-11, IRAS~4, and SS5.  The data
is shown on a relative intensity scale and are used to convey how each species
traces the high and low velocity dispersion gas.  The spectra are representative
of the FCRAO data in this study, and are shown 
at the observed resolution provided in Table~2.  Towards HH7-11 and IRAS~4 we
find clear detections of high velocity emission in CO, \HCOp , and \oHtwoO\ and
there is evidence for emission from  the low velocity dispersion quiescent gas 
in nearly every tracer.  In IRAS~4 we detect the emission of quiescent SiO
seen earlier by \citet{lefloch_sio} and \citet{codella_sio}.  In the following we 
present the distribution of these tracers separately and, more specifically,
discuss how it compares to the SWAS data.  

The emission distribution of the high velocity CO gas has been studied
in detail by \citet{ks00}, with higher resolution than obtained in this study.
However, for completeness, in Figure~4 we provide the distribution of the blue-shifted
and red-shifted (with respect to the systemic velocity) CO emission from the outflowing
gas (left-hand 
panel) along with a similar plot for \HCOp\ (right-hand panel).
In both molecules
the emission is dominated by the HH 7-11 outflow. This flow is centered on SVS~13
and fills the cavity above SVS~13 \citep{ks00}.  Weaker emission towards other 
sources (IRAS~4, IRAS~2, IRAS~7, and SVS~12) is seen as slight
extensions in the contours of this flow.  Each of the outflows are seen much more
prominently in the higher resolution IRAM data of \citet{ks00}.

Figure~5 presents integrated intensity maps for the CO isotopes, \thCO\
and \CeiO .   These maps are broadly similar to the lower resolution
\CeiO\ maps of \citet{warin_etal96}.  In particular, the cavity between
SVS~13 and SVS~12 is quite prominent.   A close comparison of the
\CeiO\ emission distribution with the \NtwoHp\ morphology presented
in Figure~1 shows that the \CeiO\ integrated intensity peaks do not
coincide with the strongest \NtwoHp\ features.  Since \NtwoHp\ closely
traces the dust emission, \CeiO\ is probably depleted
in the densest portions of the NGC 1333 core, and therefore provides an 
underestimate of the total column density.   In the following we adopt \CeiO\ as
a tracer of total column density of quiescent material and the effects of depletion
on the derived abundances is discussed in \S5.

In Figure~6 we present the observed
emission distributions for SiO (top) and \CHthreeOH\ (bottom), 
in both the high (right panels) and low (left panels) velocity 
dispersion gas.  The right-hand panels show that SiO and \CHthreeOH\ are
both detected in the IRAS~2 and IRAS~4 outflows. 
In our observations there is also weak \CHthreeOH\ emission 
in the HH7-11 outflow.
In the 50$''$ FCRAO beam we see no evidence for SiO emission in HH7-11;
however, higher resolution observations have detected SiO in this outflow
\citep{codella_sio}.  Thus, albeit with the limited statistics available in
this sample, there is some correspondence between the clearest SWAS detections of
\oHtwoO\ in molecular outflows (IRAS~2, 4, and HH7-11) and the presence
of methanol and silicon monoxide.   A similar relation between SiO and \HtwoO\
has been suggested by \citet{nisini_l1448} in a study of the L1448 outflow, 
but see also \citet{cecc_h2o}.
There does not appear to be any
\CHthreeOH\ or SiO associated with the IRAS~7 or SVS~12 outflows.

The SiO narrow component (top, left panel) emission is found predominantly
between IRAS~2, IRAS~4, and SVS~13.  The previous SiO observations of
\citet{lefloch_sio} covered an area of $\sim$2\farcm2  $\times$  4\farcm3
centered approximately between IRAS~4 and SVS~13.  The larger scale observations
presented here reveal one new clump of ambient SiO emission located 
to the north of IRAS~7.  The clump in the vicinity of IRAS~2 is
contaminated by outflow emission.  We find no evidence
for ambient SiO emission directly coincident with the 3 SWAS detections 
of quiescent water in {\em emission} (SS1, 2, 5).  Given the observed
distribution, the narrow component SiO could still
be associated with the water seen in absorption towards each of the
outflows.   
The \CHthreeOH\ narrow component (bottom, left panel) is more
distributed throughout the core, with the strongest emission clustered  
near SVS~13, IRAS~4, and IRAS~2.  The comparison of \CHthreeOH\
with SiO shows some similarities, particularly between SVS~13 and IRAS 4,
although the clump near IRAS~2 is also contaminated by outflow emission (similar to SiO).
However, there are 2 clumps of methanol emission in the northwest that are 
not observed in SiO.   Thus, there is some, but not convincing evidence of a
shock origin for ambient \CHthreeOH .
In relation to the \oHtwoO\ observations, there is no evidence for 
\CHthreeOH\ emission directly associated with SS1 and it is only weakly seen
in the cavity (SS5; Figure~3).
In addition, there is no \HtwoO\ emission towards SS3, a position coincident with
a \CHthreeOH\ clump. 

\subsection{Summary of Emission Correlations}

In this section we briefly summarize the correlations between the emission 
distributions of various tracers discussed above.
It is important to note that
molecular emission does not necessarily directly trace any given physical
variable, because a change in molecular concentration and/or cloud physical
properties (temperature, density, clump filling factor) result in changes in emission.
However, dust emission, and by association \NtwoHp , clearly delineate 
the presence of dense condensations in the NGC 1333 molecular core and 
we find no clear connection of quiescent water emission detections with these tracers.
Similarly, the mere presence of SiO in the quiescent gas is an indicator of
a shock chemistry ``legacy'' in the ambient medium and the quiescent water
emission shows no association with narrow component SiO emission.
In fact, the detections of water in absorption may trace similar gas. 
Finally, 2 of the 3 detections of quiescent water emission are towards positions with
unusually strong \thCO\ J = 5--4 emission. 

\section{ Analysis and \oHtwoO\ Abundance Determinations}

In the previous section we used the higher resolution FCRAO data to
examine the relation between the observed distribution of water
vapor to other tracers.   We have detected water vapor in 3 distinct
regimes: (1) broad emission associated with outlows, (2) narrow
emission associated with the quiescent gas, and (3) absorption
by foreground quiescent gas against the outflow emission.
In the following we make 
estimates of the water vapor abundance for each in turn. 
This analysis is complicated by the fact that
the SWAS beam is large when compared to the size scale of structures, either
outflow or clumpy substructure within NGC1333. 
In addition, to determine
abundances with observations of a single \HtwoO\ transition requires knowledge
of the density and temperature structure over a large portion of the
NGC 1333 molecular core.  These values can be estimated via multitransitional
studies, but are themselves averages of structures within a
telescopes beam and along the line of sight.
The difficulties in this process have been
discussed previously by \citet{snell_h2o} and \citet{neufeld_h2o}.   Fortunately, for
this particular core we can draw upon the wealth of data obtained previously,
and for this study, to derive simple estimates, or limits, on the average
water abundance within a SWAS beam.

\subsection{Water Abundance in the Outflows}

Based on the distribution of CO and \HCOp\ emission, the outflows, except HH7-11
are confined to regions smaller than the SWAS beam.  Thus, one uncertainty in deriving
the \oHtwoO\ abundance is 
the unknown filling factor of water emission.  The SWAS water emission could be
confined to a series of small shocked HH knots such as seen in the ISO water 
detections by  
\citet{molinari_hh711}.  Indeed the SWAS observations in Figure~1 overlap
with a number of HH objects and we have not placed on this figure the objects
with HH like nebulosities \citep{asr94, hl95}, which would increase
the correspondence.  Alternately, the emission might arise from a similar
region traced by the high velocity emission of other molecular 
tracers such as CO, CS, SiO, or \CHthreeOH . 
We treat each of these possibilities in turn.

\subsubsection{Shock Origin}

To explore the likelihood that {\em all} of the water emission seen by SWAS
arises in the shocked gas (not the extended molecular outflow seen in CO), we have
used the non-dissociative C-shock models of \citet{kn_shocks1} (hereafter KN96).
In these models the strength of 557 GHz water emission 
within a SWAS beam depends on the shock velocity,
pre-shock density, filling factor of shocked gas within the beam,
and gaseous oxygen abundance,
along with other parameters (ionization fraction, magnetic field
strength).   With suitable assumptions regarding the ionization
fraction and magnetic field strength (see KN96) we
have run a grid of shock models encompassing shock velocities
of 10, 15, and 20 \kms\ for pre-shock density of \nhtwo\ = 10$^5$
\cc .  We also examined an additional model with $v_{shock} = 15$ \kms\
and a pre-shock density of \nhtwo\ = 10$^4$ \cc .
In each model the water abundance is determined by the assumed abundance of atomic
oxygen.  The \HtwoO\ is produced solely through gas-phase neutral-neutral
reactions and not by grain mantle sputtering of \HtwoO .  This will not
affect the results, as the main dependence is on the final 
water abundance and not the exact mechanism. 

For this analysis we use corollary ISO observations
of the \oHtwoO\ \isoh2o\ 179 $\mu$m line.  Briefly, ISO has detected
179 $\mu$m water emission within a $\sim$80$''$ beam towards IRAS~4 and, with 
a lesser degree of certainty, IRAS~2 \citep{cecc_h2o}.  Observations
towards off-source positions coincident with outflow peaks found 
weak or insignificant detections \citep{giannini_h2o, maret_h2o}.
\citet{molinari_hh711} detailed ISO observations of the HH7-11 flow,
where 179 $\mu$m \oHtwoO\ emission was detected towards HH7
and SVS~13 (HH11 and HH10 in the beam).
There were no detections towards the red lobe peaks in the molecular 
outflow seen in other tracers.  
In Table~3 we provide the total integrated emission strengths for
the 557 GHz line and total fluxes for each source in the 179 $\mu$m line.   

Our shock model analysis, when compared to both ISO
and SWAS data, provides one general result. 
We are able to reproduce the observed 
179 $\mu$m flux in the weakest shock in our sample,
with a pre-shock density of \nhtwo\ = $10^4$ \cc\ and a shock velocity of 15 \kms .  
Even though the weakest shock reproduces the 179 $\mu$m emission it does not explain
the SWAS emission.   If we assume a typical HH object has a size
of 10$''$ (based on the extent of H$_2$ (1-0)S(1) emission in HH7; Molinari et al 2000),
then  a single object produces 
$\int T$dv(\oHtwoO\ \swash2o) $\sim$ 
0.38  K km s$^{-1}$.   
Adding additional shocks aids in matching the
SWAS data, but exceeds the ISO observations.  Thus, it is likely that
the SWAS emission is not from the shock itself, but rather from cold post-shock gas.
Colder temperatures allow for increases in the 557 GHz emission,
without augmenting the 179 $\mu$m flux. 
\citet{molinari_hh711} found similar problems when comparing shock models
to the ISO data, which might therefore be reconciled through the addition of 
a cold water component with elevated abundances in the high velocity gas. 
This conclusion gains additional support through the combined ISO and
SWAS study of four outflows by \citet{benedettini_h2o}, who find that the SWAS 557 GHz
emission receives a strong contribution from cold gas that is not traced by ISO. 

\subsubsection{Post-Shock Origin}

The results from shock models suggest that the 557 GHz observations could be primarily
tracing cold gas, in a manner potentially similar to that seen by other
molecules such as CO, CS, \CHthreeOH , and SiO.  
Clearly this is an 
over simplification, as some small fraction of the emission seen by SWAS must 
originate from the hot gas detected by ISO.   However, it does allow for simple 
limits to be set on the abundance.   
One complication is that
the various molecules have disparate distributions
within the SWAS beam (see Figures 4 and 5), each with its own filling factor.
To establish a lower limit to the abundance, we assume that the water is
tracing the same gas seen in low-J CO emission, which has the most widespread
emission and greatest filling factor.
An estimate of the water abundance can be derived with knowledge of the density,
temperature, and column density structure, within the outflow.  
In Table~3  we list our assumptions for these properties, which 
are justified below.

Since we have assumed that \oHtwoO\ is tracing the same gas seen in CO line
wings, we use the CO high velocity emission to estimate the total
gas column density.  The \Htwo\ column density is then derived from 
the CO column density given in Table~3 and n$_{\rm CO}$/\nhtwo\ $= 10^{-4}$.  
Density estimates in several of the NGC 1333 outflows can be found
in the literature, typically based on observations
of several transitions of CS.  Rather than adopting values from the 
literature, we have performed our own multi--transitional fit using 
the CS J=2--1, 3--2, and 5--4 data of \citet{lcl96} 
and \citet{lefloch_dust} that has been
kindly provided to us by the authors.  These data were obtained with 
the IRAM 30m with angular resolutions of 25$''$, 16$''$, and 10$''$.
Before performing the excitation analysis we convolved all data
to the lowest resolution.  Each of these maps was sampled with a 
spacing of 24$''$ on the cloud, and 12$''$ near the IRAS~2 and 4 outflows.
Thus some information is missing from the convolution for the CS J=5--4, and
to a lesser extent, the J = 3--2
observations.  To extract the outflow emission from the ambient  emission,
the CS J=2--1 data for each position were fit with a multi-component Gaussian. 
The same process was performed for the CS J=3--2 and J=5--4 data.  However,
for each position the line center velocity and velocity width were fixed
to the multi-component values derived from CS J=2--1 and only the peak intensity was allowed
to vary.  This was done both for the outflow and quiescent emission components.
The procedure produced good fits to the data for nearly all positions.

The results of this process are the integrated fluxes from the broad
emission component fit with the same linewidth for
all 3 CS transitions at every position with significant outflow J=2--1 emission.
These data were then fit using a LVG excitation code
for CS excitation in the manner described in \citet{bergin_density}.  CS emission
is generally optically thick in molecular clouds.  However in our analysis
the spread of emission in frequency due to the molecular outflow reduces the opacity. 
Our model is then simply a fit to the excitation
of the rotational ladders in statistical equilibrium.
For consistency we adopt a constant temperature for each
outflow using the values given in Table~3.   In this fashion, densities, with
constant temperature, are derived as a function of position 
where high velocity CS J=2--1 was detected.
For each outflow these densities are averaged together, weighted by the
errors, resulting in a single average density for each of the 4 outflows. 
These densities are slightly lower than quoted in the literature \citep{lcl96,
rudolph_hh711} due to our convolution to a uniform
resolution and the detailed Gaussian fitting process.

With the average physical quantities provided in Table~3 we derive a simple
estimate of the water abundance, under the assumption that the water 
emission is optically thick, but effectively thin (see Snell et al. 2000 and
Neufeld et al. 2000).   Under this assumption every collisional excitation results
in a photon that eventually escapes the cloud, which is possible when
the density is significantly below the critical density for excitation.
This condition is satisfied on average for each of the flows.   
In this case, for effectively thin emission,
the average column density of \oHtwoO\ within a SWAS beam is related 
to the 557 GHz line strength by:

\begin{equation}
\eqnum{1}
{\rm N(o-H_{2}O) = \frac{4\pi}{hc^{3}} \frac{2\nu^2 k}{C_{ul}n_{H_2}} e^{(h\nu/kT)} \int T_R dv}
\end{equation}

\noindent where C$_{ul}$ is the downwards collisional rate of \oHtwoO\ with ortho and
para-\Htwo .  
Rates of excitation of water with ortho-H$_2$ are often an order of magnitude
greater than rates for lower energy para-H$_2$ \citep{phillips_h2oxs}.  Thus assumptions
regarding the H$_2$ ortho/para ratio are required to derive water abundances and
introduce an additional source of uncertainty.
As in \citet{neufeld_h2o} we assume that all \Htwo\ is in the ground
rotational state.  
Table~3 provides the column density of \oHtwoO\ in the outflows
and the water abundance is calculated using the CO column density. 
The values derived range from 4 $\times 10^{-7}$
to 1 $\times 10^{-6}$.   The value estimated for IRAS~4
is slightly below that found by \citet{neufeld_h2o}, due to 
different assumptions regarding the density and total column density (mass) in the outflow.
Towards HH7-11 there is evidence in the high signal to noise \thCO\ J = 5--4
detection of a broad emission component below the narrow quiescent gas emission.
This broad emission component can be used to set limits on the 
column density of warm gas within the
SWAS beam and has an integrated emission of 0.43$\pm$0.04 K km/s.  Using
the physical parameters for HH7-11 in Table~3 we estimate
a CO column density of $\sim 1 \times 10^{16}$ cm$^{-2}$ is required.  
This value is quite comparable to  that estimated for ``hot'' 
CO gas by \citet{molinari_hh711} (N(CO) $\sim 10^{16}$ cm$^{-2}$).
Thus, if all the 557 GHz water emission arises from the same gas as traced 
by this component then the abundance is x(\oHtwoO) $\sim 8 \times 10^{-6}$.

Equation (1) will also hold for the \isoh2o\ transition of \oHtwoO\ observed
with ISO.  Using the densities and temperatures listed in Table~3 and the 
column density that matches the observed \swash2o\ emission we can predict the
amount of water emission at 179 $\mu$m.  These predictions are also provided 
in Table~3 and are nearly two orders of magnitude below the observed
values.  This confirms the earlier assertion that a single model cannot account for
the emission of both lines and that the 557 GHz emission likely traces
cold post-shock gas. The 179 $\mu$m emission preferentially probes either warm shocked
gas, such as seen in HH7, or, the warm envelopes surrounding protostars,
as proposed by \citet{cecc_h2o} and \citet{maret_h2o}.  

The abundances provided in Table~3 are averaged over the 
outflow and are based on a data set analyzed with
consistent assumptions for each flow.   
They represent a lower limit to the actual water abundance in the high
velocity gas,  which is due to a number of factors.
(1) If the water emission is not effectively thin then using equation (1) 
leads to an underestimate of the total column density.  
(2) We have made the conservative assumption that the \oHtwoO\ emission
traces that of CO.  Thus, the average abundances do not rule out spatial variations
in the water vapor distribution within a SWAS beam.  Such a distribution
would exist if the SWAS observations where probing the hot shocked
regions seen in ISO and any additional colder component.
This is supported by the large velocity widths of the \oHtwoO\ emission when
compared to CO (Table~3).
(3) Under the effectively thin approximation the greatest uncertainty in
the abundance lies in the determination of the density of molecular hydrogen.
We have assumed that the entirety of
the outflow traced in high velocity CO emission is 
uniformly filled with gas at a density of $\sim$10$^{5}$ \cc .  
This is likely not the case as, within a given SWAS measurement,
emission from the dense gas tracers, such as 
CS and \HCOp , are more spatially confined when compared to CO.

\subsection{Water Abundance in the Quiescent Gas}


Similar to outflow
emission, the water abundance calculation for the quiescent 
component is complicated due to structure within
the large SWAS beam.  Furthermore, the line intensity distribution of typical
tracers of the dense gas, such as dust continuum, \NtwoHp , and
CS emission do not consistently agree with the water distribution.
SS5 lies within a cavity generally (although not completely; Figure~3) 
devoid of emission from each of these tracers,
and there is little dust continuum or \NtwoHp\ emission towards SS1. 
For detections of water emission we therefore compute beam-average abundances in a fashion
similar to \S 4.1.

For the quiescent gas, the density is well below critical \citep{lcl96}, 
and the water emission is optically thick, but effectively thin.   
For the temperature we use
the observed $^{12}$CO excitation temperature towards each position determined
from the J=1--0 map of spectra convolved to a  circularized 4$'$ SWAS beam.  
This value is generally
$\sim$ 20 K. To derive the total column density, 
the \CeiO\ J=1--0 data is also convolved to 4$'$.  The total
column density is estimated from the
integrated intensity, assuming LTE at the CO excitation temperature and
a \CeiO\ fractional abundance of 1.7 $\times 10^{-7}$.

The density estimate is complicated by 
the lack of emission from tracers of the dense gas at positions 
near the quiescent \HtwoO\
detections. We therefore use the ratio of the SWAS \thCO\ J=5--4 to FCRAO \thCO\ 
J=1--0
(convolved to 4$'$) as our primary density estimate.  
Figure~7 shows the \thCO\ 5--4/1--0 ratio as a function
of density and \thCO\ opacity for a temperature of 20 K.  This ratio
is strong function of density and, with knowledge of the opacity (using
the \thCO /\CeiO\ ratio assuming $^{12}$C/$^{13}$C = 60 and
$^{16}$O/$^{18}$O = 500) and
temperature, can be used to provide an estimate of the average density within
a 4$'$ beam.  
Using these average properties, an H$_2$ ortho/para ratio thermalized at 
20 K and Eqn. (1) we derive single component
beam-average abundances which are listed in Table~4 along with the other
derived quantities.  

The water abundances are uniform at a value of $\sim 1 
\times 10^{-7}$.  The primary uncertainty in this calculation lies
in the density determination.
Here we have adopted \thCO\ as our density tracer.  \thCO\ has a low
dipole moment and is much more easily excited in the low density gas than \oHtwoO .  
Thus its emission, particularly for J=1--0, contains a contribution from
the low density gas that should not participate in the excitation
of \HtwoO\ molecules.  This would increase the density, and decrease
the water abundance.  Therefore, it is unlikely that the average
density within the beam is much lower than our adopted value and,
provided the temperature is not much higher than 20 K, our
abundances represent upper limits.   A lower abundance limit can be set by
examining the distribution of other dense gas tracers within the
SWAS beam.  The detection of ambient \oHtwoO\
emission in SVS~12 is coincident with a clump of \NtwoHp\ emission,
which traces gas with \nhtwo\ $> 10^5$ \cc .  It is clear from Figure~1
that the dense gas does not fill the SWAS beam.  If
we convolve the \NtwoHp\ FCRAO data towards this position to a resolution of 4$'$, we
find an \NtwoHp\ opacity $\sim 1$ and excitation temperature $\sim 4$ K.
From a LVG excitation model these values are roughly coincident 
at a density $\sim 10^{5}$ \cc , and 
the range of beam-average density lies between 10$^4$ -- 10$^5$ \cc .  
This limits \oHtwoO\ abundance that corresponds to these densities ranges
from $0.1 - 1 \times 10^{-7}$. 

\subsection{Absorption by Foreground Quiescent Gas}

The detections of quiescent water in absorption against outflow
emission (\S~3) can be subject to a more simplified analysis.  
As seen in Figure~1 none of the absorption features reach the
continuum-subtracted baseline.  Thus at the  1$\sigma$ level the
absorption lines are not saturated.  This could be the result
of optically thin emission or, alternately collisionally excited
water in the surrounding dense gas could fill in the absorption
trough.  
If we assume the line is resolved and unsaturated,
only a small column density of water is required
to absorb 557 GHz emission, and these small absorption features can be reproduced by
an \oHtwoO\ abundance relative to \Htwo\ of only 6 $\times 10^{-11}$.  
However, each of the absorption features is consistent, at
the 3$\sigma$ level, with saturation, in this case
the abundance is a lower limit.  Therefore
these measurements do not place stringent limits on the ambient water abundance.

\section{Discussion}

To examine the origin of the quiescent water emission in NGC 1333 
we can first compare the \oHtwoO\ abundance in NGC 1333 with those
determined towards other sources by \citet{snell_h2o} and 
\citet{bergin_b68h2o}. 
First it is clear that the quiescent ortho-water abundance in NGC 1333 is
greater than that found in cold star-less clouds. 
For instance in 
TMC-1 the \oHtwoO\ abundance relative to \Htwo\ is $< 7 \times 10^{-8}$
\citep{snell_h2o}.   
More stringent limits ($x$(\oHtwoO ) $< 1-7 \times 10^{-9}$)
have been set towards B68 and $\rho$ Oph Core D \citep{bergin_b68h2o}. 
These two cores have well described physical properties providing
more accurate abundance determinations. 
In NGC 1333 the lower limit to the quiescent water abundance (1 $\times 10^{-8}$) 
is well above the upper limits found in B68 and $\rho$ Oph D. 
Similarly, on average, the quiescent water abundance in NGC 1333
appears to be higher than derived towards denser and more massive star-forming
cores such as OMC-1 or M17 where abundances range from $0.1 - 1 \times 10^{-8}$
\citep{snell_omc1, snell_m17}.  
However, it is comparable to values derived in low density clouds
along the line of sight towards Sgr B2 and Sgr A$^*$ \citep{cernicharo_h2o, neufeld_h2o,
moneti_h2o}.  

One concern is the use of \CeiO\ as a column density tracer in our
calculation, and in the previous abundance estimates.  In NGC 1333 \CeiO\
appears to be depleted and the primary effect
of this depletion would be to raise the \oHtwoO\ abundance. 
If \CeiO\ depletion is more significant in NGC 1333 than in B68 or
$\rho $ Oph D then the abundance disparity could be reconciled.  However,
this is unlikely as the colder clouds should have more significant \CeiO\
depletion, and hence higher abundances.  
In sum it appears that the water abundance in NGC 1333 is enhanced, certainly
when compared to colder pre-stellar molecular cores.  Herschel observations of ortho 
and para forms of water, and additional transitions, in the ambient gas will 
provide the data to confirm or refute this result.

Given the widespread nature of outflow activity, the quiescent water abundance  
enhancement in NGC 1333 could be the result of gas exposure to numerous shocks.  
Indeed in \S 4.1 we found evidence that the water in the high velocity gas is likely 
cold and therefore has begun the process of enriching the surrounding gas with 
cold water vapor.
However, through our analysis of other tracers of shock activity (SiO and \CHthreeOH ),
we find little or no evidence to support this assertion.  All positions where 
quiescent water emission is detected (SS1, SS2-SVS12, SS5) show no quiescent 
SiO emission and only 1 position coincides with strong \CHthreeOH\ emission. 
Thus, despite the appeal, this hypothesis has no support from the molecular 
data.

However, quiescent \HtwoO\ emission is correlated with 
detections of unusually strong \thCO\ J = 5--4 emission.  
This emission is in close proximity
to the nearby B stars (BD +30.549 and SVS~3) 
that power the reflection nebula.  This region has been identified
as a PDR through observations of infrared emission arising from PAHs 
\citep{joblin_1333, uchida_iso}.   
PDR's are characterized in terms
of the enhancement of the FUV field, G$_0$, measured in units of the equivalent
flux of 1.6 $\times 10^{-3}$ ergs cm$^{-2}$ s$^{-1}$ determined by Habing (1968)
for the interstellar medium. 
The stellar luminosities for these stars are given by \citet{harvey_ir}
(BD +30.549: 400 L$_{\odot}$; SVS~3: 360 L$_{\odot}$), 
with these values we estimate G$_0$ $\sim 400$ towards the two northernmost
SWAS survey positions (SS1, SS2-SVS12) and G$_0$ $\sim 100$ towards SS5.
These FUV enhancements provide additional heating that excites the high-J lines
of \thCO\ that we observe.  The strongest \thCO\ J = 5--4 emission arises in lower
density gas closest to the interface and will have the largest FUV enhancements
(e.g. SS1, SS2).  However, \thCO\ J = 5--4 
emission is also observed throughout the NGC 1333 cloud (e.g. Figure~2).   
Towards SS6-8 (HH7-11, IRAS~2, IRAS~4) this emission is coincident with
dense condensations traced with \NtwoHp , thus the higher density, 
and embedded protostars, provide added excitation to account 
for any reduction in heating from the external FUV field.  

Given the evidence for enhanced FUV radiation we have explored whether a PDR could
give rise to the observed \oHtwoO\ emission, as suggested by \citet{spaans_h2o}.
The PDR calculations presented here are from the model of Tielens \& 
Hollenbach (1985), updated by Kaufman et al. (1999). The model 
self-consistently calculates the temperature and equilibrium chemical structure
of gas illuminated by far-ultraviolet radiation from nearby stellar 
sources, given the FUV field strength \G0\ and the gas density, and
outputs the expected line intensities from atomic and
molecular species. In the outer, atomic regions ($A_V\lesssim 3-5$),
cooling is primarily through far-infrared emission from CII and OI, while
further in ($A_V\gtrsim $ 5) cooling by molecular species such as CO and
$^{13}$CO becomes important.

In our plane-parallel geometry for the PDR model we fix the depth to 11 mag by the observed 
\CeiO\ column density (scaled by the \CeiO\ abundance).  This
is clearly an approximation as the true extinction between the external
sources and the water emitting gas is unknown.
We assume \nhtwo\ $= 2 \times 10^{4}$ \cc\ based on the multi-transitional
\thCO\ analysis.   
To predict the water abundance profile we use the 
time-dependent gas-grain chemical model
of \citet{bergin_ic5146} using the same density and the gas temperature
profile as in the PDR calculations.  We adopt the new H$_3$O$^{+}$ dissociative
recombination branching ratios of \citet{neau_h3op}.
Models with these parameters were run for values
of G$_0$ = 1, 100, and 400, predicting the fluxes from \thCO\ J = 5--4 and
\oHtwoO\ \swash2o.
Figure 8 presents the model gas temperature and water abundance profile,
along with the integrated emission as a function of extinction.
Table~6 provides the average abundances and integrated fluxes from the model.

The \HtwoO\ abundance is reduced at low extinction due to 
photodestruction and rises towards higher extinction as the UV photons are
absorbed by dust grains.  Because the gas is at relatively low density we
have assumed that the water molecules have not yet had sufficient time to
freeze out as in dense cores.
For \nhtwo\ = 2 $\times 10^{4}$ \cc , models with \G0\ = 1 do not produce significant 
\oHtwoO\ \swash2o\ or \thCO\ J = 5--4 emission.  
When the FUV field is raised, the line 
fluxes increase, and for \G0\ $\ge 100 - 400$ the model reproduces both the
quiescent \HtwoO\ and high-J \thCO\ observations. 
In cases with enhanced UV fields
the average water abundance is $x$(\oHtwoO) $\sim 1.5 \times 10^{-7}$ (assuming
ortho-\HtwoO /para-\HtwoO = 3). This is close to the derived quiescent water
abundance suggesting that simple gas phase chemistry in a PDR may be enough
to account for the SWAS observations. 

In all, the details are certainly more complex
than included in the model presented here.  
However, additional support for the presence
of an active PDR is found in the ISO detections of [C II] 158 $\mu$m and
[O I] 63 and 145 $\mu$m emission towards the HH7-11 flow.  
\citet{molinari_hh711} provide fluxes of 21.3 $\times 10^{-20}$ W/cm$^{-2}$
([C II] 158 $\mu$m), 180 $\times 10^{-20}$ W/cm$^{-2}$ ([O I] 63 $\mu$m),
and 8.0 $\times 10^{-20}$ W/cm$^{-2}$ ([O I] 145 $\mu$m) towards SVS13.
Assuming a higher density of 10$^{5}$ cm$^{-3}$ for this
gas (\S 4), the PDR models can reproduce the flux and flux ratios of these
lines provided G$_0 \sim 50 - 100$. From  
the luminosities of the external sources and the relative distances to SVS13
we estimate G$_0 \sim 70$, consistent with the PDR model.
Thus the model and available data are both
suggestive of the importance of the PDR in NGC 1333 for creating the 
observed atomic and quiescent water emission.

The excitation of water in a PDR may therefore
account for the water emission seen by SWAS towards other massive star forming
regions, such as OMC-1 and M17.  This may also account for the 
remarkable similarity in the spectral line profiles between \oHtwoO\ \swash2o\
and CO J = 2--1 noted by \citet{ashby_profile}.
We stress that this is not in conflict with the oxygen freeze-out models
of \citet{bergin_impl}, because freeze-out is still required to account for the lack of
a stronger signature in the water emission from gas that is 
clearly associated with regions of higher density, \nhtwo\ $> 10^{5}$ \cc\
(i.e. the OMC-1 core).  Here the
water would be depleted in the dense regions, but not in the low density outer
layers with longer depletion timescales and higher temperatures.

\section{Conclusions}

We have presented the results of a biased survey for \swash2o\ emission
from \oHtwoO\ in the core of the NGC 1333 molecular cloud using
the Submillimeter Wave Astronomy Satellite.  This survey is biased in
the sense that known centers of activity were centered within
the 3\farcm3 $\times$ 4\farcm5 SWAS beam. 
These data are supplemented by
an unbiased FCRAO survey in the 
J=1--0 transitions of $^{12}$CO, $^{13}$CO, \CeiO , \NtwoHp , \HCOp ,
and the J=2--1 transition of SiO.  Each of these transitions were mapped,
with beam sampling, over an area that encompasses all SWAS observations.
The primary results are listed below.

(1) We report the detection of broad \oHtwoO\ emission in the IRAS~2, IRAS~7, and
HH7-11 outflows.  The detection of \swash2o\ emission in the IRAS~4 flow has
been reported previously \citep{neufeld_h2o}.   In each case the emission
is accompanied by narrow absorption features at the quiescent cloud velocity.
Towards 3 positions we detect narrow ($\Delta v < 3$ \kms ) emission
clearly associated with the quiescent gas.  
 

(2) Using these data sets we
derive beam-averaged abundances for outflow emission in NGC 1333.
In the outflows the water abundance relative to \Htwo\ is $>$ 10$^{-6}$.
These abundances are averaged
within the large SWAS beam and, hence, we have no information on structures
smaller than $4'$.   However, with this caveat in mind, our results confirm
the enhancement of \oHtwoO\ in the NGC~1333 molecular outflows.
A combination of published observations, and those reported here,
find broad emission in SiO and \CHthreeOH\ in the HH7-11, IRAS~2, and IRAS~4 outflows.
These same sources are the strongest centers of \oHtwoO\ outflow activity 
detected by SWAS.  Although the sample
is quite limited, this is suggestive of a potential correlation between regions with 
SiO and \CHthreeOH\ abundance enhancements and detectable water emission. 
Models of non-dissociative (C-type) shocks are used to constrain
the origin of the 557 GHz outflow emission seen by SWAS and 
179 $\mu$m emission detected by ISO.   These models suggest that the
water abundance is enhanced within the shocks, but fail to reproduce
the water emission unless the
majority of the \swash2o\ emission arises from cold post-shock gas within a portion of the
extended outflows.  

(3) In the quiescent gas beam-averaged abundances lie between $0.1 - 1 \times 10^{-7}$.   
The quiescent water emission is correlated with unusually strong
\thCO\ J = 5--4 emission and is found near the
NGC 1333 reflection nebula and PDR.  
Through a PDR model we find that
the local enhancements of the FUV field increase the gas temperature in relatively low
density gas (\nhtwo\ $\sim 10^4$ \cc ) allowing for simple gas phase chemistry 
(with slightly lower gas phase abundances
due to photodissociation) to  account  observed fluxes in both \thCO\ J = 5--4 and
\oHtwoO\ \swash2o.  
These results are consistent the results of 
\citet{spaans_h2o} who examined water emission in S140. 
Thus the excitation
of water in low density gas, that is undepleted due to longer depletion timescales
and exposed to local enhancements of the FUV field, may account for the extended
557 GHz water emission found towards a variety of molecular sources.
However, depletion of oxygen in the form of water is still required for regions 
with significant amounts of dense gas along the line of sight.

\acknowledgements

This work was supported by NASA's SWAS Grant NAS5-30702.
The Five College Radio Astronomy Observatory is operated with the
support of the National Science Foundation under grant AST01-00793.
EAB is grateful to helpful discussions with C. Ceccarelli and S.T. Megeath,
and to B. Lefloch and A. Castets for providing the multitransitional CS data.
We also are grateful to the referee for providing a thorough and thoughful review.

\begin{deluxetable}{crrrcc}
\tablenum{1}
\tablecolumns{6}
\tablewidth{5.0in}
\tablecaption{SWAS Survey Sample Positions}
\tablehead{
\colhead{SWAS Survey No.} &
\colhead{$\Delta \alpha (')$} &
\colhead{$\Delta \delta (')$} &
\colhead{Source\tablenotemark{a}} &
\colhead{Time (hrs)} &
\colhead{rms (mK)\tablenotemark{b}} 
}
\startdata
SS1 & $-$0.8 & 7.2 & \nodata & 32.8 & 8 \\ 
SS2 & $-$1.5 & 5.0  & SVS 12 & 35.6 & 11 \\ 
SS3 & $-$5.6 & 3.2 & \nodata & 30.0 & 10 \\ 
SS4 & 1.5 & 2.2 &  IRAS 7 & 61.0 & 7 \\ 
SS5 & $-$2.2 & 2.2  & \nodata & 57.5 & 8 \\ 
SS6 & 0.0 & 0.0   & SVS 13 & 29.6 & 11 \\ 
SS7 & $-$1.6 & $-$1.6 & IRAS 2 & 29.0 & 12 \\ 
SS8 & 1.5  & $-$2.1 & IRAS 4 & 37.3 & 9 \\ 
SS9 & $-$4.2 & $-$4.3 & \nodata & 26.2 & 10 \\ 
\enddata
\tablenotetext{a}{Only associations with well studied IRAS or SVS sources are noted.}
\tablenotetext{b}{Typical rms in 557 GHz band, system temperatures were $\sim 2100$ K
with minimal scatter around this value.}
\end{deluxetable}

\begin{deluxetable}{llrrrcrrrr}
\tabletypesize{\scriptsize}
\tablewidth{7.0in}
\tablenum{2}
\tablecolumns{10}
\tablecaption{Observed Transitions and Telescope Parameters}
\tablehead{
\colhead{Molecule} &
\colhead{Transition } &
\colhead{$\nu$(GHz)} &
\colhead{$E_u$ (K)} &
\colhead{Telescope} & 
\colhead{$\Delta v_{res}$ (km s$^{-1}$)} & 
\colhead{$\theta_{MB}$} &
\colhead{$\eta_{MB}$} &
\colhead{t$_{int}$ (m)} &
\colhead{$\sigma (T_A^*)$ (mK)} \\ 
}
\startdata
H$_2$O\tablenotemark{a}    & $J = 1_{10} \rightarrow 1_{01}$ & 556.93600 & 61.0\tablenotemark{b} & SWAS & 
1.00 & 3\farcm3 $\times$ 4\farcm5 & 0.90 & \nodata & \nodata \\ 
$^{13}$CO\tablenotemark{a}  & $J = 5 \rightarrow 4$ & 550.92630 & 79.3 & SWAS & 1.00 & 3\farcm3 $\times$ 4\farcm5 & 0.90 & \nodata & \nodata \\ 
$^{12}$CO & $J = 1 \rightarrow 0$ & 115.27120 & 5.5  & FCRAO & 0.52 & 45$''$ & 0.45 &
6.5 & 200 \\ 
$^{13}$CO & $J = 1 \rightarrow 0$ & 110.20137 & 5.3 & FCRAO & 0.21 & 48$''$ & 0.50 & 1.3
& 300 \\ 
C$^{18}$O & $J = 1 \rightarrow 0$ & 109.78218 & 5.3 & FCRAO & 0.43 & 47$''$ & 0.50 & 6.8
& 90 \\ 
CH$_3$OH & $J = 2_0 \rightarrow 1_0$A$^+$ & 96.74142 & 7.0 & FCRAO & 0.48 & 54$''$  & 0.50  & 16.0 & 30   \\ 
N$_2$H$^+$ & $J = 1 \rightarrow 0$ & 93.17378 & 4.5 & FCRAO & 0.50 & 56$''$ & 0.50 & 4.3
& 70 \\ 
HCO$^+$ & $J = 1 \rightarrow 0$ & 89.18855 & 4.3  & FCRAO & 0.53 & 58$''$ & 0.45 & 11.3
& 50\\ 
SiO      & $J = 2 \rightarrow 1$ & 86.84696 & 6.3 & FCRAO & 0.54 & 60$''$  & 0.45 & 18.0
& 30\\ 
\enddata
\tablenotetext{a}{Integration time and sensitivities given in Table 1.}
\tablenotetext{b}{Transition is 27 K above o-H$_2$O ground state.}
\end{deluxetable}

\begin{deluxetable}{lllll}
\tablewidth{6.0in}
\tablenum{3}
\tablecolumns{6}
\tablecaption{Data, Estimated Source Properties, and Results -- NGC 1333 Outflows} 
\tablehead{

\colhead{} &
\colhead{IRAS 4} &
\colhead{IRAS 2} &
\colhead{HH7-11} &
\colhead{IRAS 7} \\

\colhead{} &
\colhead{(1.5,--2.1)} &
\colhead{(--1.6,--1.6)} &
\colhead{(0.0,0.0)} &
\colhead{(1.5,2.2)} \\ 
}
\startdata
\underline{o-H$_2$O 1$_{10}-1_{01}$ (SWAS):} & & & & \\
$\int\;T_A^*dv$ (K km/s) & 1.27$\pm$0.06 & 1.89$\pm$0.08
& 1.60$\pm$0.06 & 0.51$\pm$0.03  \\
$\Delta v$ (km s$^{-1}$) & 24.4$\pm$0.8 & 24.7$\pm$1.3 & 27.8$\pm$1.5 
& 8.5$\pm$0.8\\
& & & &  \\
\underline{o-H$_2$O 2$_{12}-1_{01}$ (ISO):} & & &  \\
$\int\;S_{\nu}d\nu$ (W cm$^{-2}$)\tablenotemark{a} 
& 2.8 $\times 10^{-19}$ & 1.6 $\times 10^{-19}$ & 1.5 $\times 10^{-19}$ & \nodata \\
& & & & \\
\underline{CO 1--0 (FCRAO-conv. 4$'$):}  & & & &  \\
 $\int\;T_A^*dv$ (K km/s) & 8.9$\pm$0.3 & 14.7$\pm$0.3
& 21.5$\pm$0.3 & 8.9$\pm$0.4 \\
$\Delta v$ (km s$^{-1}$) & 11.4$\pm$0.3 & 11.1$\pm$0.2 & 11.7$\pm$0.1 & 12.0$\pm$0.5 
\\\hline
\underline{Estimated Properties and Results:} & & & & \\
n(H$_2$) (cm$^{-3}$) & $1 \times 10^5$ & $3 \times 10^5$ & $5 \times 10^4$ 
& $6 \times 10^4$\\
Temperature (K) & 75\tablenotemark{b} & 40\tablenotemark{c} & 50\tablenotemark{d} & 40\tablenotemark{e}  \\ 
$\int\;S_{\nu}d\nu$ (W cm$^{-2}$) [179 $\mu$m]\tablenotemark{f} & 3.5 $\times 10^{-21}$ & 2.8 $\times 
10^{-21}$ & 2.3  $\times 10^{-21}$ & 1.1  $\times 10^{-21}$ \\ 
N(o-H$_2$O) (cm$^{-2}$) & 3 $\times 10^{14}$ & 3 $\times 10^{14}$ & 1 $\times 10^{15}$
& 6 $\times 10^{14}$  \\
N(CO) (cm$^{-2}$)\tablenotemark{g} & 8 $\times 10^{16}$ & 7 $\times 10^{16}$ & 1 $\times 10^{17}$
& 4 $\times 10^{16}$  \\
x(o-H$_2$O) & 4 $\times 10^{-7}$ & 4 $\times 10^{-7}$ & 1 $\times 10^{-6}$
& 1 $\times 10^{-6}$  \\
\enddata
\tablenotetext{a}{Total fluxes from outflow derived using data obtained from the ISO data 
archive using ISAP version 2.0a and LWS calibration version 8.}
\tablenotetext{b}{Estimated by Blake et al. (1995) from CS and H$_2$CO.}
\tablenotetext{c}{Adopt value used by \citep{lcl96} in their analysis of
multitransitional CS data.}
\tablenotetext{d}{\citet{rudolph_hh711} estimate T = 10 -- 40 K on the basis of
VLA NH$_3$ observations, while \citet{codella_sio} derive 50 K from SiO observations.
We adopt the latter value.}
\tablenotetext{e}{No previous estimates in the literature.}
\tablenotetext{f}{Predicted \oHtwoO\ 179 $\mu$m emission from single component 
(constant density and temperature) model that reproduces the 557 GHz emission (\S
4.1.2).}
\tablenotetext{g}{Estimated by convolving FCRAO data to 4$'$.  To separate
outflow and quiescent gas emission the convolved spectra were fit with multi-component
Gaussians.  Column density is derived assuming LTE at the temperature listed here.}
\end{deluxetable}

\begin{deluxetable}{llll}
\tablewidth{5.0in}
\tablenum{4}
\tablecolumns{4}
\tablecaption{Data, Estimated Source Properties, and Results --\hspace{1in} NGC 1333
Ambient Gas } 
\tablehead{

\colhead{} &
\colhead{SS1} &
\colhead{SVS 12} &
\colhead{SS5}  \\

\colhead{} &
\colhead{(--0.8,7.2)} &
\colhead{(--1.5,5.0)} &
\colhead{(--2.1,2.2)} 
}
\startdata
\underline{o-H$_2$O 1$_{10}-1_{01}$ (SWAS):} & & &  \\
$\int\;T_A^*dv$ (K km/s) & 0.19$\pm$0.02 & 0.24$\pm$0.03
& 0.19$\pm$0.02 \\
$v$ (km s$^{-1}$) & 6.4$\pm$0.2 & 6.9$\pm$0.2 & 7.3$\pm$0.2 \\
$\Delta v$ (km s$^{-1}$) & 3.7$\pm$0.6 & 3.0$\pm$0.4 & 2.8$\pm$0.5 \\
& & & \\
\underline{C$^{18}$O 1--0 (FCRAO-conv. 4$'$):}  & & & \\
 $\int\;T_A^*dv$ (K km/s) & 1.67$\pm$0.04 & 1.71$\pm$0.04
& 1.35$\pm$0.04 \\
$v$ (km s$^{-1}$) & 7.9$\pm$0.1 & 8.0$\pm$0.1 & 8.1$\pm$0.1 \\
$\Delta v$ (km s$^{-1}$) & 1.9$\pm$0.1 & 2.1$\pm$0.1 & 2.2$\pm$0.1 \\\hline
\underline{Estimated Properties and Results:} & & & \\
n(H$_2$) (cm$^{-3}$) & $2 \times 10^4$ & $1 \times 10^4$ & $1 \times 10^4$ \\
Temperature (K) & 20 & 20 & 20 \\ 
N(o-H$_2$O) (cm$^{-2}$) & 3 $\times 10^{15}$ & 3 $\times 10^{15}$ & 2 $\times 10^{15}$ \\
N(C$^{18}$O) (cm$^{-2}$) & 4 $\times 10^{15}$ & 5 $\times 10^{15}$ & 3 $\times 10^{16}$\\
x(o-H$_2$O) & 1 $\times 10^{-7}$ & 1 $\times 10^{-7}$ & 8 $\times 10^{-8}$ \\
\enddata
\end{deluxetable}

\begin{deluxetable}{crrrccc}
\tablenum{5}
\tablecolumns{7}
\tablewidth{6.7in}
\tablecaption{$^{13}$CO J = 5--4 Line Parameters}
\tablehead{
\colhead{SWAS Survey No.} &
\colhead{$\Delta \alpha (')$} &
\colhead{$\Delta \delta (')$} &
\colhead{Source} &
\colhead{$\int$T$_A^*$d$v$ (K km s$^{-1}$)} &
\colhead{$v$ (km s$^{-1}$)} &
\colhead{$\Delta v$ (km s$^{-1}$)} 
}
\startdata
SS1 & $-$0.8 & 7.2 & \nodata  &  4.34$\pm$0.06 & 7.9$\pm$0.1 & 2.3$\pm$0.1\\ 
SS2 & $-$1.5 & 5.0  & SVS 12  &  4.03$\pm$0.06 & 8.0$\pm$0.1 & 2.2$\pm$0.1\\ 
SS3 & $-$5.6 & 3.2 & \nodata  &  0.26$\pm$0.05 & 8.5$\pm$0.1 & 1.9$\pm$0.1\\ 
SS4 & 1.5 & 2.2 &  IRAS 7     &  1.89$\pm$0.06 & 8.5$\pm$0.1 & 2.2$\pm$0.1\\ 
SS5 & $-$2.2 & 2.2  & \nodata & 1.44$\pm$0.05 & 8.3$\pm$0.1 & 2.5$\pm$0.1\\ 
SS6 & 0.0 & 0.0   & SVS 13    & 2.42$\pm$0.05 & 8.4$\pm$0.1 & 2.4$\pm$0.1\\ 
SS7 & $-$1.6 & $-$1.6 & IRAS 2 & 1.86$\pm$0.05 & 8.1$\pm$0.1 & 2.4$\pm$0.1\\ 
SS8 & 1.5  & $-$2.1 & IRAS 4 & 1.10$\pm$0.05 & 8.0$\pm$0.1 & 2.6$\pm$0.1\\ 
SS9 & $-$4.2 & $-$4.3 & \nodata & 0.28$\pm$0.06 & 7.8$\pm$0.1 & 3.1$\pm$0.1\\ 
\enddata
\end{deluxetable}

\begin{deluxetable}{lrrrrrrr}
\tablewidth{7.0in}
\tablenum{6}
\tablecolumns{8}
\tablecaption{ PDR Model Results for NGC1333} 
\tablehead{

\colhead{} &
\colhead{} &
\colhead{} &
\colhead{} &
\multicolumn{2}{c}{$^{13}$CO J = 5--4 (K km/s)} &
\multicolumn{2}{c}{o-H$_2$O 1$_{10}-1_{01}$ (K km/s)}  \\

\colhead{Position} &
\colhead{G$_0$} &
\colhead{$<x$(H$_2$O)$>$} &
\colhead{$x$(H$_2$O)$_{max}$} &
\colhead{Model} &
\colhead{Obs.} &
\colhead{Model} &
\colhead{Obs.} 
}
\startdata
SS1  & 400 & 2.0 $\times 10^{-7}$ & 4 $\times 10^{-7}$ & 2.25 & 4.34$\pm$0.06 & 0.25 & 0.19$\pm$0.02
\\
SS2(SVS12) & 400 & 2.0 $\times 10^{-7}$ & 4 $\times 10^{-7}$ & 2.25 & 4.03$\pm$0.06  & 0.25 & 0.24$\pm$0.02 \\
SS5 & 100 & 1.8 $\times 10^{-7}$ & 4 $\times 10^{-7}$ & 1.35 & 1.44$\pm 0.05$ &
0.17 & 0.19$\pm$0.02 \\
&  1 & 3.6 $\times 10^{-7}$ & 4 $\times 10^{-7}$ & 0.01 & \nodata & 0.03 & \nodata \\
\enddata
\end{deluxetable}

\clearpage

\begin{figure}
\plotone{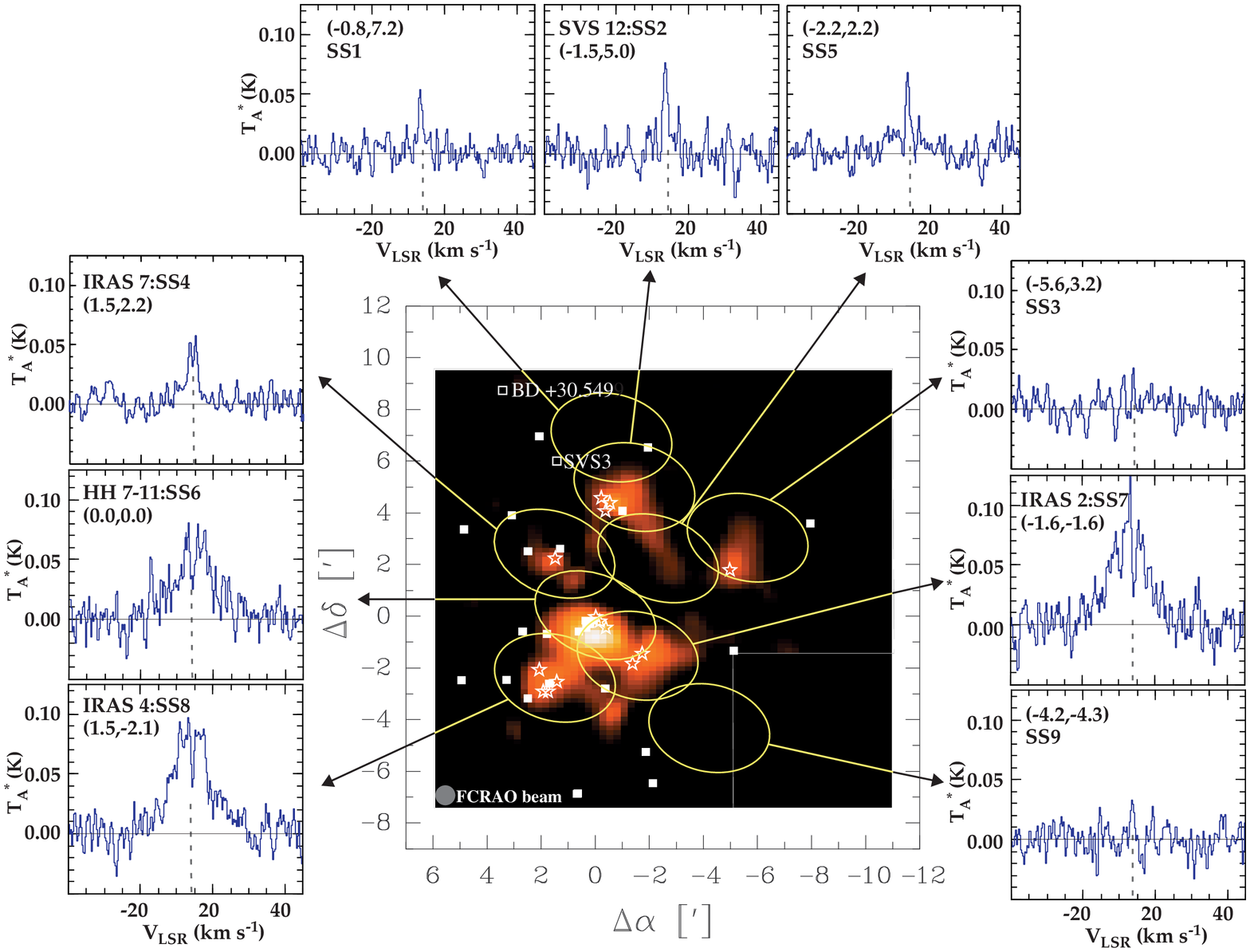}
\figurenum{1}
\figcaption{
Spectra of the 1$_{10} - 1_{01}$ 556.936 GHz transition 
of ortho-water vapor from the various positions in NGC 1333.  
Each spectra is labeled with any prominent source that is centered or
within the beam, along with the SWAS survey position number (Table 1).
Placements of the elliptical
SWAS beam at the observed rotation angle are shown superposed on
a map of integrated N$_2$H$^+$ (J=1--0) emission.  
The scale of the N$_2$H$^+$ emission ranges
from 1 K km s$^{-1}$ to 8 K s$^{-1}$ and is integrated over all hyperfine components.
The dashed line in each spectra is the velocity of the ambient gas as determined
by Gaussian fits to the  N$_2$H$^+$ spectra (convolved to a circularized 4$'$ 
SWAS beam).  The stars show the location of IRAS sources and SVS~13 (0.0,0.0), 
while the solid squares are positions of known
HH objects (see text for references).  The open squares denote the positions of two
stars associated with reflection nebulosity, BD +30.549 and SVS~3.
Labels for other well known sources are provided in Figure 4. 
The map is reference to the position of SVS~13.
}
\end{figure}

\begin{figure}
\plotone{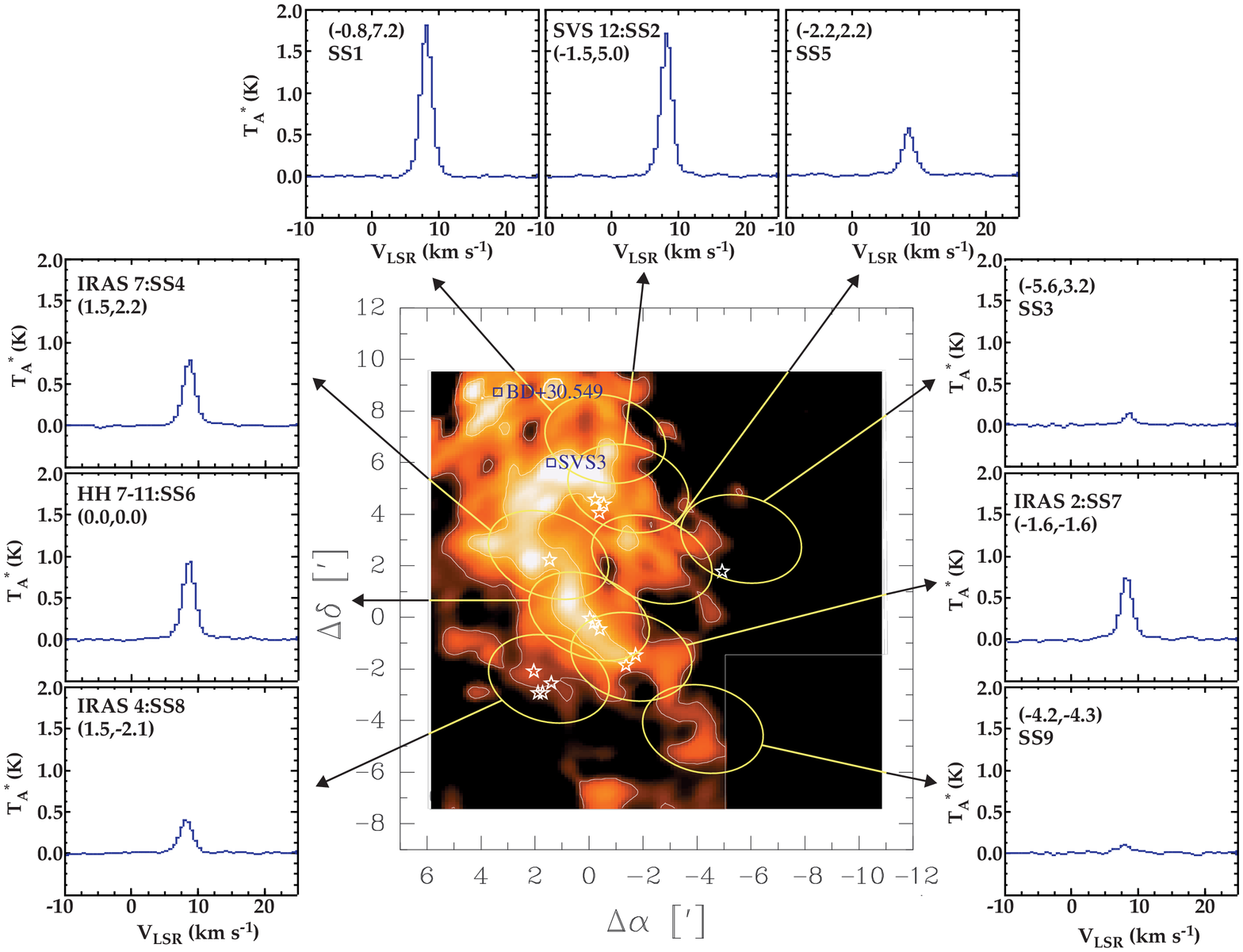}
\figcaption{
Spectra of the J = 5--4 transition of $^{13}$CO
from the various positions in NGC 1333.  
Each spectra is labeled with any prominent source that is centered or
within the beam, along with the SWAS survey position number (Table 1).
Placements of the elliptical
SWAS beam at the observed rotation angle are shown superposed on
a map of integrated $^{13}$CO (J=1--0) emission.  
The scale of the  $^{13}$CO (J=1--0) emission ranges
from 8 K km s$^{-1}$ to 22 K s$^{-1}$ and is clipped to emphasize the strongest
regions of emission. 
The stars show the location of IRAS sources and SVS~13 (0.0,0.0).
The open squares denote the positions of two
stars associated with reflection nebulosity, BD +30.549 and SVS~3.
Labels for other well known sources are provided in Figure 4. 
The map is reference to the position of SVS~13.
}
\end{figure}

\begin{figure}
\vspace{8.0in}
\includegraphics{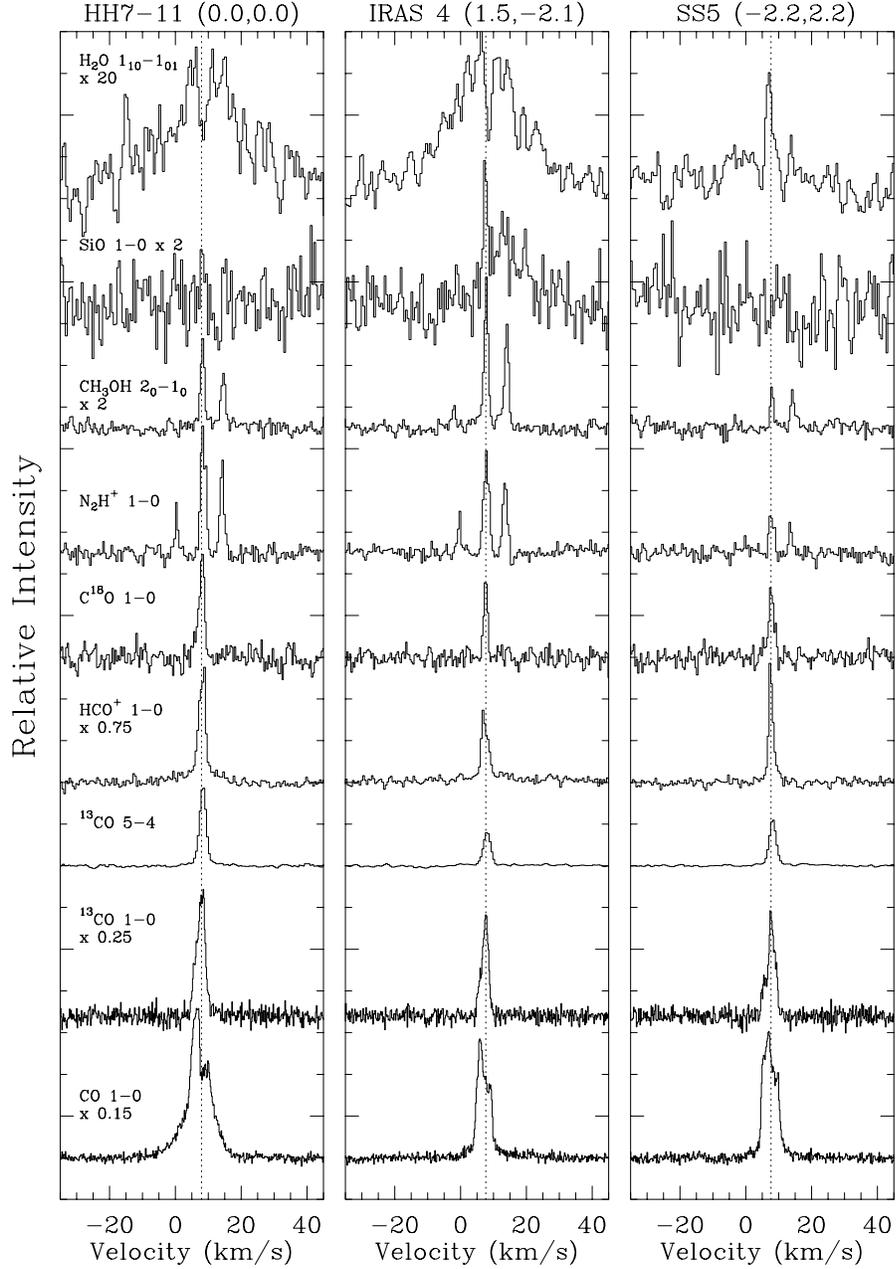}
\figcaption{Representative spectra of each molecule and transition in this study
towards positions coincident with SVS~13 (HH7-11), IRAS~4, and SS5.
The spectra are given at the observed resolution (see Table 2).  The y-axis is
relative intensity, but for scale towards SVS~13  the HCO$^{+}$ J=1--0 peak intensity is
T$_A^* =$ 1.83 K, IRAS~4 T$_A^* =$ 1.14 K, SS5 T$_A^* =$ 1.35 K.   
The vertical line is the 
cloud velocity determined via Gaussian fits to the N$_2$H$^+$ emission.
}
\end{figure}

\begin{figure}
\vspace{2.0in}
\includegraphics{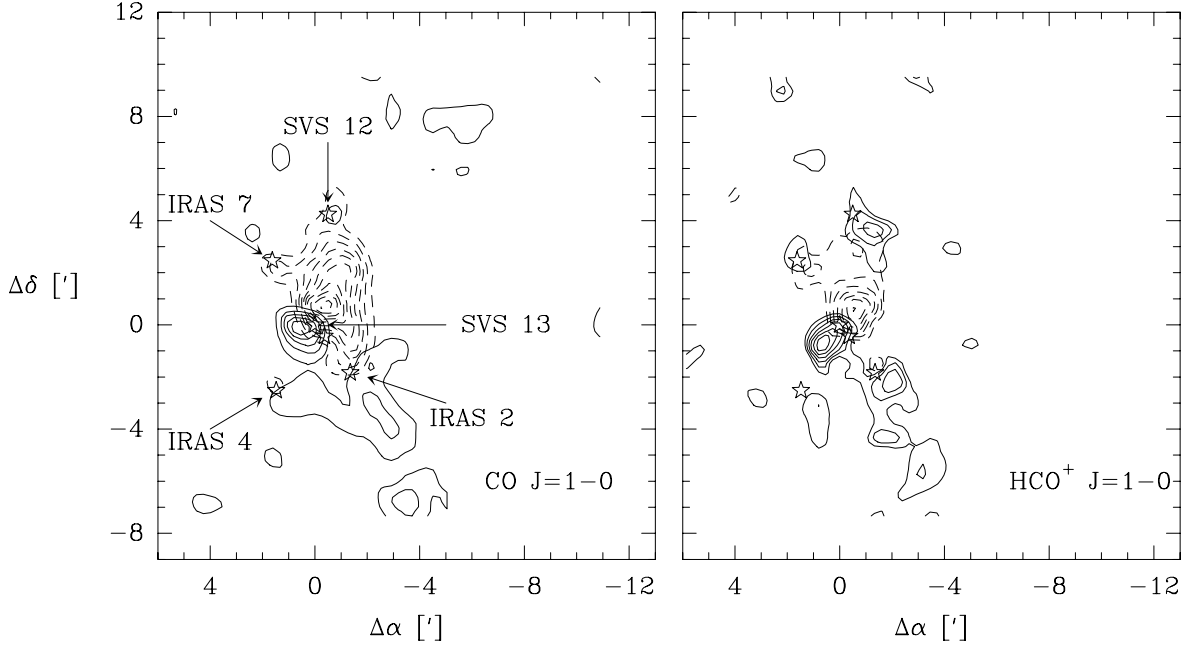}
\figcaption{
Left: Distribution of high velocity CO (J=1--0) emission in NGC 1333.  
Blue-shifted($-$20 --
4 km s$^{-1}$) is shown as solid contours and red-shifted(11 - 30 km s$^{-1}$) as
dashed contours.  Levels begin at 3 K km s$^{-1}$ (3$\sigma$) and
step at 3 K km s$^{-1}$.  Locations and names of various well known sources are
provided.
Right:  Distribution of high velocity HCO$^{+}$ (J=1--0) emission.
Blue-shifted(-8 -- 5 km s$^{-1}$) is given as solid contours and
red-shifted(11 -- 20 km s$^{-1}$) as dashed contours.  Levels start at
0.6 K km s$^{-1}$ (3$\sigma$) and step at 0.2 K km s$^{-1}$. 
The maps are referenced to the position of SVS~13.}
\end{figure}

\clearpage
\begin{figure}
\vspace{2.0in}
\includegraphics{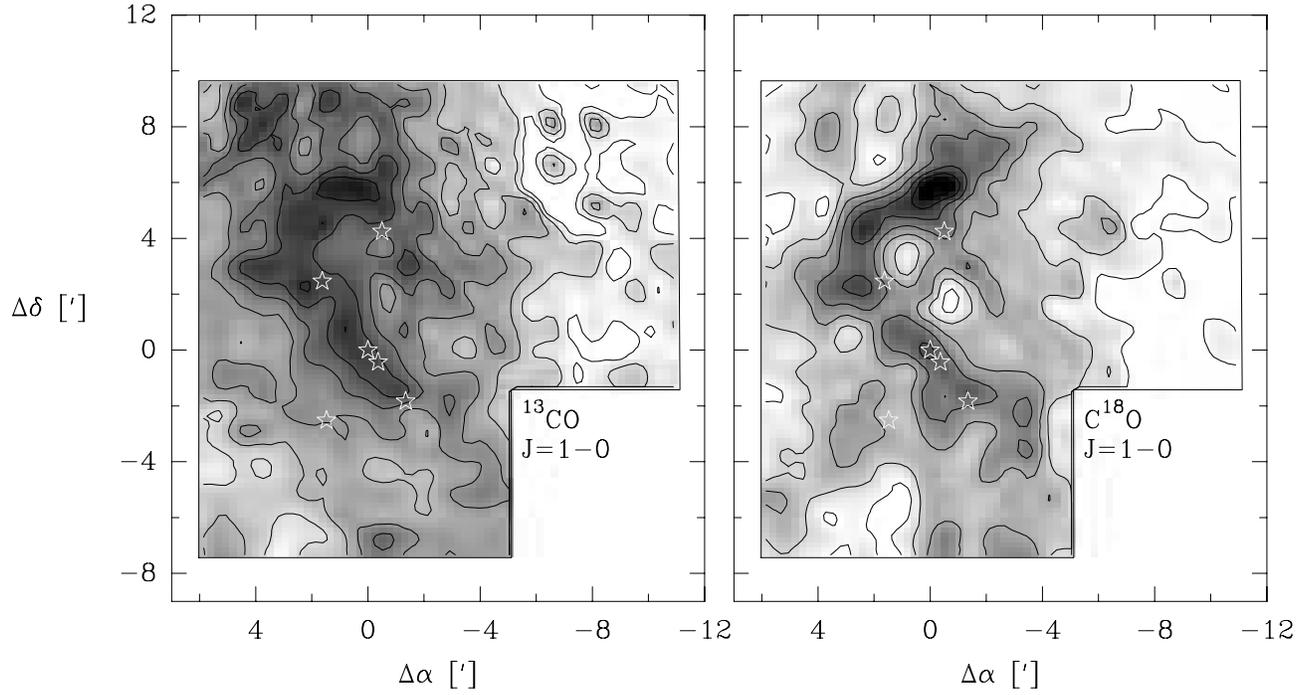}
\figcaption{
Integrated emission maps of $^{13}$CO J=1--0 (left) and C$^{18}$O J=1 -- 0 (right).
For both panels the levels begin and step at the 3$\sigma$ value, 
3.0 K km s$^{-1}$ for $^{13}$CO and 0.5 K km s$^{-1}$ for C$^{18}$O.
The maps are referenced to the position of SVS~13.
}
\end{figure}

\clearpage
\begin{figure}
\vspace{6.0in}
\includegraphics{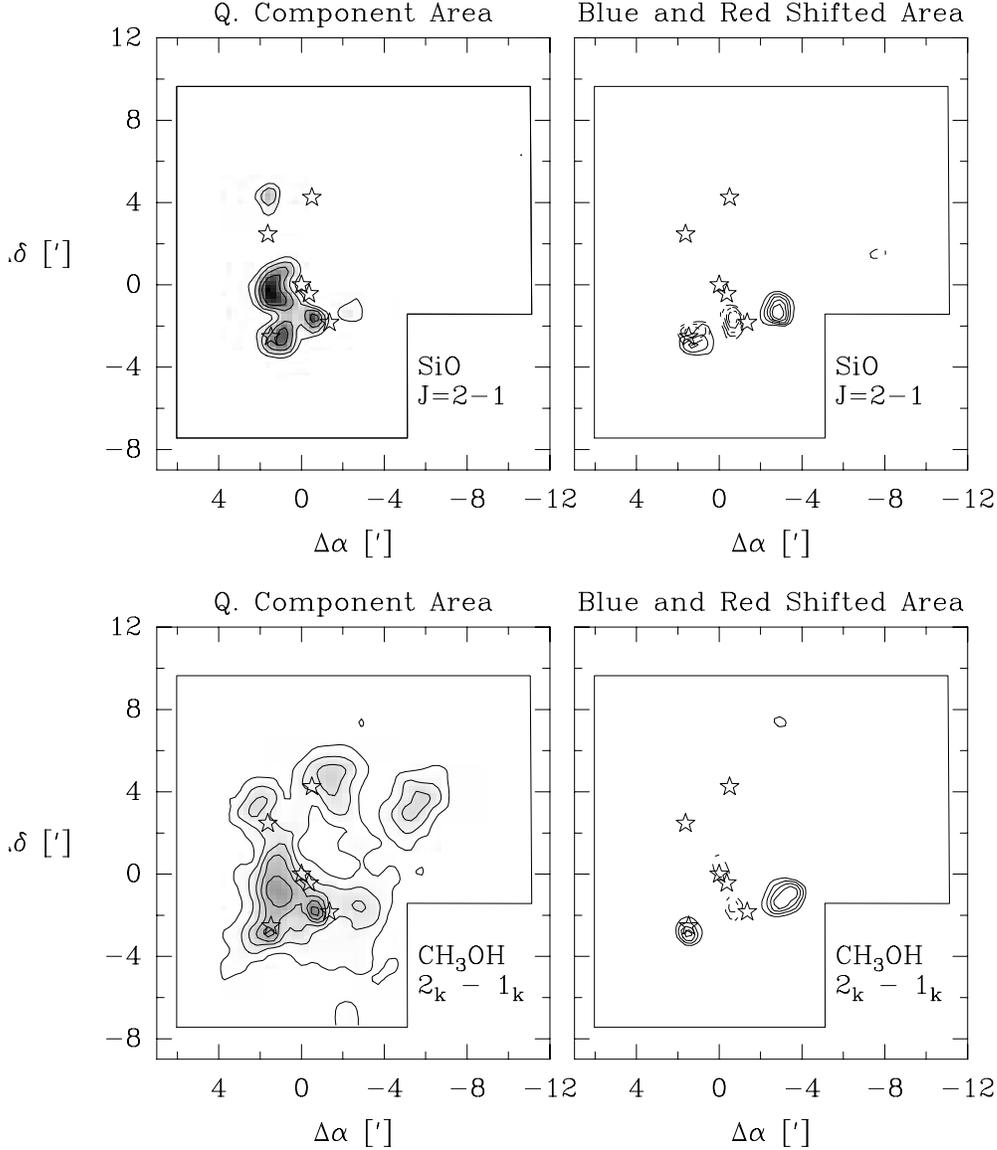}
\figcaption{
Top Panels [Left]: Integrated emission map of thermal SiO J=2--1 in
the ambient(6.3 -- 9.5 km s$^{-1}$) gas.  Contour levels begin at the 3$\sigma$
level of 0.15 K km s$^{-1}$ and step at 0.1 K km s$^{-1}$.
[Right]: Distribution of high velocity SiO J=2--1 emission.  
Blue-shifted(--7 -- 6 km s$^{-1}$) emission is given as solid contours and
red-shifted(9 -- 26 km s$^{-1}$) as dashed.  For both blue- and
red-shifted emission the contours begin at the 3$\sigma$ level of 0.3 K km s$^{-1}$
and step at 0.2 K km s$^{-1}$.
Bottom Panels [Left]: Integrated emission of CH$_3$OH J$_k = 2_k - 1_k$ 
($k$ = 0-E, 0-A$^+$, and --1-E) in the ambient(-2.6 -- 0.6, 6.3 -- 10.0, 12.6 -- 16.0
km s$^{-1}$)
gas.  Contour levels begin and step at the 3$\sigma$ level of 0.3 K km s$^{-1}$.
[Right:] Distribution of high velocity CH$_3$OH J$_k = 2_k - 1_k$ emission.
Blue-shifted(-9 -- 6  km s$^{-1}$) emission is given as solid contours and
red-shifted(16 -- 32  km s$^{-1}$) as dashed contours.  The velocity range is chosen to
avoid confusion with the strongest $k$ components.  The blue-shifted
emission contains minimal contamination from the $2_0 - 1_0$E component in the ambient
gas.  Contour levels for
both begin at the 3$\sigma$ level of 0.2 K km s$^{-1}$  
and step at 0.3 K km s$^{-1}$.
All maps are referenced to the position of SVS~13.
}
\end{figure}

\clearpage
\begin{figure}
\plotone{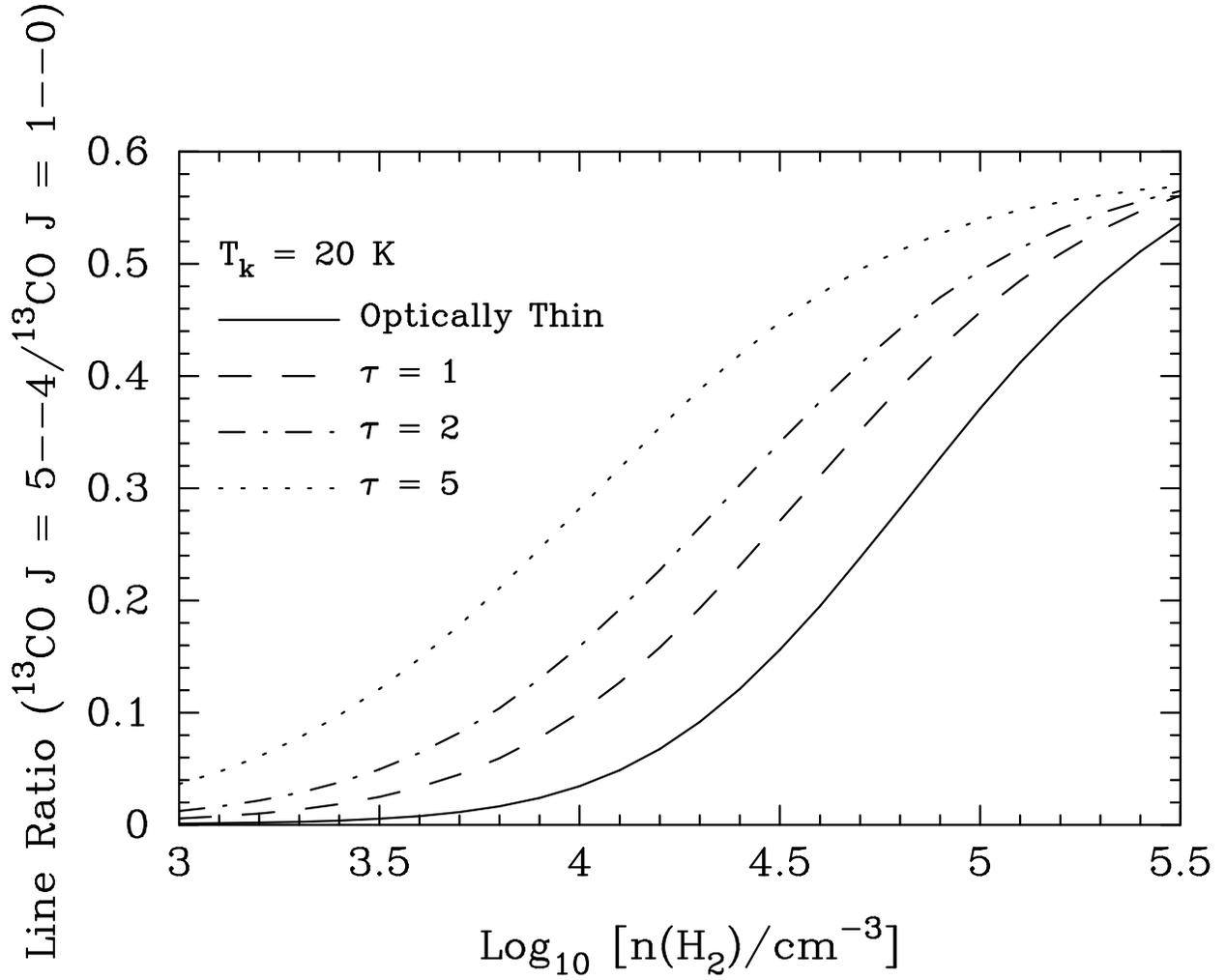}
\figcaption{$^{13}$CO J=5--4/J=1--0 line intensity ratio from a large velocity
gradient excitation
model plotted as a function of density and \thCO\ J = 1--0 opacity (column density) for
T = 20 K. }
\end{figure}

\begin{figure}
\plotone{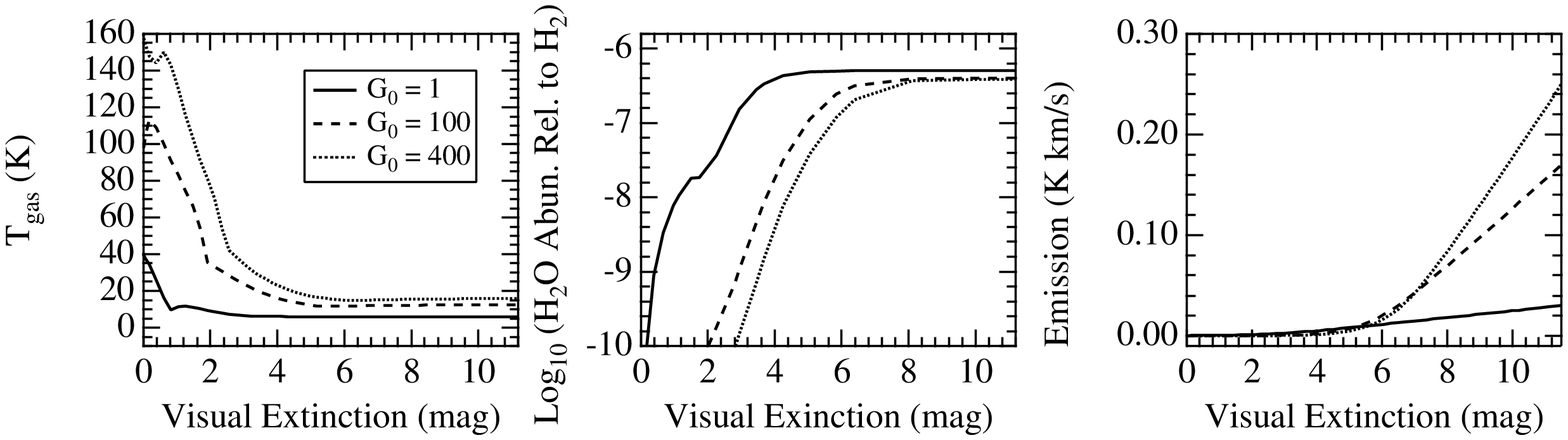}
\figcaption{
[Left:] Gas temperature as a function of cloud depth (visual extinction) for
three different PDR models.  [Middle:] Water abundance relative to H$_2$ as a function of
cloud depth for same models. [Right:] Plot of integrated emission in K km/s 
as a function of visual extinction.
}
\end{figure}


\begin{thebibliography}{}




\bibitem[Ashby et al.(2000)]{ashby_profile} Ashby, M.~L.~N.~et al.\ 2000, \apjl,
539, L115

\bibitem[Aspin, Sandell, \& Russell(1994)]{asr94} Aspin, C., 
Sandell, G., \& Russell, A.~P.~G.\ 1994, \aaps, 106, 165 

\bibitem[Aspin \& Sandell(1997)]{as97} Aspin, C.~\& Sandell, 
G.\ 1997, \mnras, 289, 1 

\bibitem[Bachiller(1996)]{bachiller_araa} Bachiller, R.\ 1996, \araa, 34, 111

\bibitem[Bachiller \& Cernicharo(1990)]{bc90} Bachiller, R.~\&
Cernicharo, J.\ 1990, \aap, 239, 276

\bibitem[Bally, Devine, \& Reipurth(1996)]{bdr96} Bally, J., Devine, D.,
\& Reipurth, B.\ 1996, \apjl, 473, L49

\bibitem[Benedettini et al.(2002)]{benedettini_h2o} Benedettini, M., Viti, S.,
Giannini, T., Nisini, B., Goldsmith, P.F., \& Saraceno, P. 2002, A\&A, in press

\bibitem[Bergin \& Snell (2002)]{bergin_b68h2o} Bergin, E.A. \& Snell, R.L. 2002, ApJL, 
submitted 

\bibitem[Bergin et al.(2001)]{bergin_ic5146} Bergin, E.~A., Ciardi, D.~R., Lada,
C.~J., Alves, J., \& Lada, E.~A.\ 2001, \apj, 557, 209

\bibitem[Bergin et al.(2000)]{bergin_impl} Bergin, E.~A.~et al.\ 2000, \apjl,
539, L129

\bibitem[Bergin, Snell, \& Goldsmith(1996)]{bergin_density} Bergin, E.~A., Snell,
R.~L., \& Goldsmith, P.~F.\ 1996, \apj, 460, 343


\bibitem[Bergin, Melnick, \& Neufeld (1998)]{bmn98} Bergin, E.~A., 
Melnick, G.~J., Neufeld, D.~A.\ 1998, \apj, 499, 777

\bibitem[Blake et al.(1995)]{blake_iras4} Blake, G.~A., Sandell, G., van
Dishoeck, E.~F., Groesbeck, T.~D., Mundy, L.~G., \& Aspin, C.\ 1995, \apj, 441, 689

\bibitem[Caselli, Hartquist, \& Havnes(1997)]{caselli_sio} Caselli, P.,
Hartquist, T.~W., \& Havnes, O.\ 1997, \aap, 322, 296

\bibitem[Ceccarelli et al.(1999)]{cecc_h2o} Ceccarelli, C.~et al.\ 1999,
\aap, 342, L21


\bibitem[Cernicharo et al.(1997)]{cernicharo_h2o} Cernicharo, J.~et 
al.\ 1997, \aap, 323, L25 

\bibitem[Cernis(1990)]{cernis90} Cernis, K.\ 1990, \apss, 166, 315

\bibitem[Charnley, Rodgers, \& Ehrenfreund(2001)]{charnley_swasiso} Charnley, S.~B., R
odgers, S.~D., \& Ehrenfreund, P.\ 2001, \aap, 378, 1024. 

\bibitem[Codella, Bachiller, \& Reipurth(1999)]{codella_sio} Codella, C.,
Bachiller, R., \& Reipurth, B.\ 1999, \aap, 343, 585

\bibitem[Di Francesco et al.(2001)]{difrancesco_iras4} Di Francesco, J., 
Myers, P.~C., Wilner, D.~J., Ohashi, N., \& Mardones, D.\ 2001, \apj, 562, 
770 



\bibitem[Fraser, Collings, McCoustra, \& 
Williams(2001)]{fraser_h2olab} Fraser, H.~J., Collings, M.~P., 
McCoustra, M.~R.~S., \& Williams, D.~A.\ 2001, \mnras, 327, 1165 

\bibitem[Giannini, Nisini, \& Lorenzetti(2001)]{giannini_h2o} Giannini, T.,
Nisini, B., \& Lorenzetti, D.\ 2001, \apj, 555, 40


\bibitem[Graff \& Dalgarno(1987)]{graff_oxy} Graff, M.~M.~\& 
Dalgarno, A.\ 1987, \apj, 317, 432 

\bibitem[Habing(1968)]{habing_uv} Habing, H.~J.\ 1968, \bain, 19, 
421 

\bibitem[Harvey, Wilking, \& Joy(1984)]{harvey_ir} Harvey, P.~M., 
Wilking, B.~A., \& Joy, M.\ 1984, \apj, 278, 156 


\bibitem[Harwit, Neufeld, Melnick, \& Kaufman(1998)]{harwit_h2o} Harwit, M.,
Neufeld, D.~A., Melnick, G.~J., \& Kaufman, M.~J.\ 1998, \apjl, 497, L105


\bibitem[Hodapp \& Ladd(1995)]{hl95} Hodapp, K.~\& Ladd, E.~F.\ 1995,
\apj, 453, 715

\bibitem[Joblin, Tielens, Geballe, \& Wooden(1996)]{joblin_1333} 
Joblin, C., Tielens, A.~G.~G.~M., Geballe, T.~R., \& Wooden, D.~H.\ 1996, 
\apjl, 460, L119 


\bibitem[Kaufman \& Neufeld(1996)]{kn_shocks1} Kaufman, M.~J.~\& Neufeld,
D.~A.\ 1996, \apj, 456, 611

\bibitem[Kaufman, Wolfire, Hollenbach, \& 
Luhman(1999)]{kaufman_pdf} Kaufman, M.~J., Wolfire, M.~G., 
Hollenbach, D.~J., \& Luhman, M.~L.\ 1999, \apj, 527, 795 

\bibitem[Knee \& Sandell (2000)]{ks00} Knee, L.~B.~G.~\& 
Sandell, G.\ 2000, \aap, 361, 671 

\bibitem[Lada, Alves, \& Lada(1996)]{lal96} Lada, C.~J., 
Alves, J., \& Lada, E.~A.\ 1996, \aj, 111, 1964 

\bibitem[Langer, Castets, \& Lefloch(1996)]{lcl96} Langer, 
W.~D., Castets, A., \& Lefloch, B.\ 1996, \apjl, 471, L111 

\bibitem[Lefloch et al.(1998)]{lefloch_sio} Lefloch,
B., Castets, A., Cernicharo, J., \& Loinard, L.\ 1998, \apjl, 504, L109

\bibitem[Lefloch et al.(1998)]{lefloch_dust} Lefloch, B., Castets, 
A., Cernicharo, J., Langer, W.~D., \& Zylka, R.\ 1998, \aap, 334, 269 


\bibitem[Liseau et al.(1996)]{liseau_hh54} Liseau, R.~et al.\ 1996, 
\aap, 315, L181 


\bibitem[Maret et al.(2002)]{maret_h2o} 
Maret S., Ceccarelli C., Caux E., Tielens A.G.G.M., Castets A., Parise B. 2002,
A\&A, submitted



\bibitem[May et al.(2000)]{may_sputter} May, P.~W., Pineau des For{\^ e}ts, G.,
Flower, D.~R., Field, D., Allan, N.~L., \& Purton, J.~A.\ 2000, \mnras, 318, 809

\bibitem[Melnick et al.(2000a)]{melnick_bnkl} Melnick, G.~J.~et al.\ 
2000a, \apjl, 539, L87 

\bibitem[Melnick et al.(2000b)]{melnick_swas} Melnick, G.~J.~et al.\ 2000b, \apjl,
539, L77

\bibitem[Molinari et al.(2000)]{molinari_hh711} Molinari, S.~et al.\ 2000, \apj,
538, 698

\bibitem[Moneti, Cernicharo, \& Pardo(2001)]{moneti_h2o} Moneti, 
A., Cernicharo, J.~;, \& Pardo, J.~R.~;.\ 2001, \apjl, 549, L203 

\bibitem[Neau et al.(2000)]{neau_h3op} Neau, A.~et al.\ 2000, 
\jcp, 113, 1762 

\bibitem[Neufeld et al.(2000a)]{neufeld_h2o} Neufeld, D.~A.~et al.\ 2000a, \apjl,
539, L107

\bibitem[Neufeld et al.(2000b)]{neufeld_sgrb2} Neufeld, D.~A.~et al.\ 
2000b, \apjl, 539, L111 

\bibitem[Nisini et al.(1999)]{nisini_h2o} Nisini, B.~et al.\ 1999, \aap, 350,
529

\bibitem[Nisini et al. (2000)]{nisini_l1448} Nisini, B., Benedettini, M., 
Giannini, T., Codella, C., Lorenzetti, D., Di Giorgio, A.M., \& Richer,
J. S.\ 2000, \aap , 360, 297


\bibitem[Phillips, Maluendes, \& Green(1996)]{phillips_h2oxs} 
Phillips, T.~R., Maluendes, S., \& Green, S.\ 1996, \apjs, 107, 467 



\bibitem[Reipurth (1999)]{reipurth_hhcat} Reipurth, B. 1999, {\em A general
catalog of Herbig-Haro objects}, 2. edition, http://casa.colorado.edu/hhcat

\bibitem[Rudolph et al.(2001)]{rudolph_hh711} Rudolph, A.~L., Bachiller, R., Rieu,
N.~Q., Van Trung, D., Palmer, P., \& Welch, W.~J.\ 2001, \apj, 558, 204

\bibitem[Sandell \& Knee (2001)]{sk01} Sandell, G.~;.~\& 
Knee, L.~B.~G.\ 2001, \apjl, 546, L49 

\bibitem[Sandell et al.(1994)]{sandell_iras2} Sandell, G., Knee, L.~B.~G., Aspin,
C., Robson, I.~E., \& Russell, A.~P.~G.\ 1994, \aap, 285, L1

\bibitem[Schilke, Walmsley, Pineau Des Forets, \& Flower(1997)]{schilke_sio}
Schilke, P., Walmsley, C.~M., Pineau Des Forets, G., \& Flower, D.~R.\ 1997, \aap, 321,
293

\bibitem[Snell et al.(2000a)]{snell_h2o} Snell, R.~L.~et al.\ 2000a, \apjl, 539,
L101

\bibitem[Snell et al.(2000b)]{snell_omc1} Snell, R.~L.~et al.\ 2000b, \apjl, 539, L93. 

\bibitem[Snell et al.(2000c)]{snell_m17} Snell, R.~L.~et al.\ 2000c, \apjl, 539, L97. 


\bibitem[Snell \& Edwards(1981)]{snell_edwards} Snell, R.~L.~\& 
Edwards, S.\ 1981, \apj, 251, 103 

\bibitem[Spaans \& van Dishoeck(2001)]{spaans_h2o} Spaans, M.~\& van Dishoeck, E.~F.\
2001, \apjl, 548, L217. 

\bibitem[Tafalla et al.(2002)]{tafalla_depletion} Tafalla, M.~et al.\ 2002, \apj, 
in press

\bibitem[Tielens \& Hagen(1982)]{tielens_hagen} Tielens, 
A.~G.~G.~M.~\& Hagen, W.\ 1982, \aap, 114, 245 

\bibitem[Tielens \& Hollenbach(1985)]{th_pdr} Tielens, 
A.~G.~G.~M.~\& Hollenbach, D.\ 1985, \apj, 291, 722 

\bibitem[Uchida, Sellgren, Werner, \& 
Houdashelt(2000)]{uchida_iso} Uchida, K.~I., Sellgren, K., Werner, 
M.~W., \& Houdashelt, M.~L.\ 2000, \apj, 530, 817 

\bibitem[Viti et al.(2001)]{viti_h2o} Viti, S., Roueff, E., Hartquist, T.~W., Pineau des
For{\^ e}ts, G., \& Williams, D.~A.\ 2001, \aap, 370, 557. 

\bibitem[Wagner \& Graff(1987)]{wagner_oxy} Wagner, A.~F.~\& 
Graff, M.~M.\ 1987, \apj, 317, 423 

\bibitem[Warin et al.(1996)]{warin_etal96} Warin, S., Castets, A., Langer, W.~D.,
Wilson, R.~W., \& Pagani, L.\ 1996, \aap, 306, 935

\bibitem[Wright et al.(2000)]{wright_h2o} Wright, C.~M., van Dishoeck, E.~F.,
Black, J.~H., Feuchtgruber, H., Cernicharo, J., Gonz{\' a}lez-Alfonso, E., \& de Graauw,
T.\ 2000, \aap, 358, 689

\bibitem[Ziurys, Friberg, \& Irvine(1989)]{ziurys_sio} Ziurys, L.~M., Friberg,
P., \& Irvine, W.~M.\ 1989, \apj, 343, 201




\end{thebibliography}
\end{document}